\documentclass[pra,twoside,amsfonts,amssymb,amsmath,floatfix,letterpaper,twocolumn]{revtex4}
\usepackage{amsmath}
\usepackage{amsfonts}
\usepackage{amssymb}
\usepackage{graphicx}
\usepackage{graphicx,array}
\usepackage{dcolumn}
\usepackage{bm}

\newcommand{\ket}[1]{\ensuremath{\vert #1 \rangle}}

\begin{document}

\title{Bloch theory of  entangled photon generation in non-linear photonic crystals}

\author{William T.M. Irvine}
\email[corresponding author: ]{william@physics.ucsb.edu}
\affiliation{Department of Physics, University of California, Santa Barbara, CA 93106, USA}

\author{Michiel J.A. de Dood}
\altaffiliation{Present address: Huygens Laboratory, Leiden University, PO Box 9504, 2300 RA Leiden, The Netherlands}
\affiliation{Department of Physics, University of California, Santa Barbara, CA 93106, USA}

\author{Dirk Bouwmeester}
\affiliation{Department of Physics, University of California, Santa Barbara, CA 93106, USA}

\date{\today}

\begin{abstract}
We present a quantum mechanical description of parametric
down-conversion and phase-matching of Bloch-waves in non-linear
photonic crystals. We discuss the theory in one-dimensional Bragg
structures giving a recipe for calculating the down-converted
emission strength and direction. We exemplify the discussion by
making explicit analytical predictions for the emission amplitude
and direction from a one-dimensional  structure that consists  of
alternating layers of  Al$_{0.4}$Ga$_{0.6}$As and Air. We show
that the emission is suitable for the extraction of
polarization-entangled photons.
\end{abstract}

\maketitle

\section{Introduction}

Entangled photon pairs play  a central role in both fundamental tests of quantum mechanics and in the implementation of quantum information theory protocols~\cite{dik}.
They are an appealing resource for quantum communication since they propagate easily over long distances with relatively little interaction with the environment. Furthermore
 they have  been proposed as a resource for all-optical quantum computation~\cite{klm}.

A popular method to produce entangled photons is by parametric
down-conversion in naturally birefringent non-linear crystals. In
this process `pump' photons entering a crystal decay or
`down-convert' into photon pairs. The role of the non-linearity is
to mediate the interaction whereas the role of the birefringence
is to ensure that the process is phase-matched, i.e. that the
amplitudes for the down-conversion process at different points in
the crystal  constructively interfere.   If a particular emission
geometry can be achieved, photons emerging in a specific pair of
directions will be entangled in polarization~\cite{paul}.

In a recent paper, the authors  proposed the use of non-linear
{\it photonic} crystals as a source of polarization-entangled
photons~\cite{us}. The scheme  harnesses the higher $\chi^{(2)}$
nonlinearity of semiconductor materials (e.g. $\chi^{(2)}_{\mathrm{GaAs}}=200 \mathrm{pm/V}$~\cite{chi2} c.f. $\chi^{(2)}_{\mathrm{BBO}}=2.2 \mathrm{pm/V}$~\cite{chi2bb0}) to mediate the
down-conversion  and proposes to use the photonic crystal geometry
to phase-match the emission. The scheme has the potential of both
increasing the efficiency of the process and providing an
entangled-photon  source that is  more amenable to integration on
optical chips. 
The scheme is fundamentally different from schemes for quasi phase-matching in periodically-poled materials, where only the $\chi^{(2)}$ is modulated periodically.

One-dimensional photonic crystals had been considered before for classical frequency 
conversion in the limit that the optical wavelength is much larger than the periodicity~\cite{formbi,deros,earlyexp}. 
Two-dimensional non-linear photonic crystals  have also been
considered for classical frequency
conversion~\cite{earlyth}.  Following the appearance of our proposal,
Reference~\cite{sergienk} used semi-classical coupled mode theory
to calculate co-linear frequency down-conversion efficiency in
one-dimensional structures using numerical calculations.

In this article we present the theory that underlies our proposal.
We use the Bloch-wave formalism to discuss the quantum mechanical down-conversion
process in non-linear photonic crystals and show how to calculate
the strength and direction of the down-converted emission. The theory works for
photonic crystals of all dimensions. We apply the theory to
one-dimensional Bragg structures, 
performing  calculations of the down-conversion emission analytically. By plotting the calculated emission from a
structure that consists  of alternating layers of
Al$_{0.4}$Ga$_{0.6}$As and Air, we show explicitly that entangled photon
pairs can be generated in a realistic  structure.


The present article is structured as follows:
sections~\ref{sec:pcwaves} and~\ref{sec:quantiz}  discuss  Bloch
waves and their quantization in linear photonic crystals following
essentially the work of Caticha and Caticha~\cite{caticha}. In
section~\ref{sec:pm} we derive an expression for the interaction
Hamiltonian and  the phase-matching function that govern the
down-conversion process in non-linear photonic crystals.  
 In section~\ref{sec:yariv} we summarize the discussion of
Bloch waves in Bragg structures by Yariv and Yeh~\cite{yariv}
extending it to obtain expressions for the Bloch-wave Fourier
components. Section~\ref{sec:fff} consists of a detailed
discussion of the phase-matching problem in Bragg structures,
giving a recipe for establishing  the emission amplitude and
direction from a given structure. The discussion centres on the application of our method to an example  Al$_{0.4}$Ga$_{0.6}$As/Air structure.    Finally we conclude and discuss
extensions of the present work in section~\ref{sec:concl}.

\section{Wave propagation inside a linear photonic crystal}
\label{sec:pcwaves}

A photonic crystal is a material with a periodic variation in the index of refraction or dielectric constant. A periodic dielectric constant $\epsilon(\mathbf{r})=\epsilon(\mathbf{r}+\bm{\Lambda})$, is seen as a periodic potential by the electric field. The eigen-solutions of Maxwell's equations in a medium with periodic dielectric must therefore take the form of Bloch waves.  We can thus write the following  expression for the four-vector potential inside the medium:
\begin{equation}
A^\mu_{\mathbf{K},\lambda}({\bf r},t)=e^{-i(\mathbf{K\cdot r}-\omega_{\mathbf{K},\lambda} t)}\sum_\mathbf{G}\tilde{\alpha}_{\mathbf{K},\lambda}^\mu(\mathbf{G})e^{i {\bf G\cdot r}}.
\label{eq:amu}
\end{equation}
$\mathbf{K}$ represents the Bloch momentum (and will be taken to
lie within the first Brillouin zone), the index $\lambda$ runs
over the two polarizations and the various branches of the
dispersion relation, $\mathbf{G}$ represents  a reciprocal lattice
vector and $\mu$ runs from 0 to 3. The bold face represents a
Cartesian vector  and only the real part of $A^\mu$ is of physical
significance. The sum defines a periodic envelope-field
$\alpha_{\mathbf{K},\lambda}^\mu(\mathbf{r})=\alpha_{\mathbf{K},\lambda}^\mu(\mathbf{r}+\bm{\Lambda})$
with Fourier coefficients
$\tilde{\alpha}_{\mathbf{K},\lambda}(\mathbf{G})$.  There are two natural generalizations
of the Coulomb gauge: $\epsilon(r) \nabla\cdot\mathbf{A}=0$ and
$\nabla\cdot(\epsilon(r)\mathbf{A})=0$ that reduce to the uniform
dielectric Coulomb gauge $\nabla\cdot\mathbf{A}=0$. 
In the gauge
$\epsilon(r)\nabla\cdot\mathbf{A}=0$, Maxwell's equations are
given by~\cite{caticha}:
\begin{eqnarray}
\partial_tA^0&=&\epsilon^{-1}(r)\nabla^2\mathbf{A}-\partial_t^2\mathbf{A}  \label{eq:pot1}\\
\nabla^2 A^0&=&-\epsilon^{-1}(r)\nabla \epsilon(r)\cdot(\partial_t \mathbf{A}+\nabla A^0)
\label{eq:pot2}
\end{eqnarray}
Solving these equations with the ansatz~(Eq. \ref{eq:amu}) amounts to finding the dispersion relation between $\omega$ and $\bm{K}$ and an expression for $\alpha_{\mathbf{K},\lambda}^\mu(\mathbf{r})$ or its Fourier coefficients $\tilde{\alpha}_{\mathbf{K},\lambda}^\mu(\bm{G})$. The most striking consequence of the presence of a periodic potential is the formation of frequency regions, known as `stop-bands' in which no propagating solution exists. Close to these regions, the dispersion is strongly modified and $\alpha_{\mathbf{K},\lambda}(\bm{r})$ takes the form of a standing wave. Away from these regions $\alpha_{\mathbf{K},\lambda}(\bm{r})$ recovers its plane-wave form but the dispersion relation can differ considerably from that in a uniform dielectric in a way that can be tuned by changing the geometry and strength of the periodic modulation.
The tunable dispersion relation will play a crucial role in this paper.


The electric and magnetic field can be derived   from the vector potential in the usual way:
\begin{eqnarray*}
\mathbf{E}_{\mathbf{K},\lambda} &=& - \nabla \phi_{\mathbf{K},\lambda}-\partial_t \mathbf{A}_{\mathbf{K},\lambda}\\
\mathbf{B}_{\mathbf{K},\lambda} &=& \nabla \times \mathbf{A}_{K,\lambda}
\end{eqnarray*}
Where $\phi_{\mathbf{K},\lambda}=A^0_{\mathbf{K},\lambda}$ and $\mathbf{A}=(A^1_{\mathbf{K},\lambda},A^2_{\mathbf{K},\lambda},A^3_{\mathbf{K},\lambda})$, giving:
\begin{eqnarray}
\mathbf{E}_{\mathbf{K},\lambda}({\bf r},t)&=&e^{-i(\mathbf{K\cdot r}-\omega_{\mathbf{K},\lambda} t)}\sum_\mathbf{G}\tilde{\bm{\varepsilon}}_{\mathbf{K},\lambda}(\mathbf{G})e^{i {\bf G\cdot r}} \label{eq:ef}\\
\mathbf{B}_{\mathbf{K},\lambda}({\bf r},t)&=&e^{-i(\mathbf{K\cdot r}-\omega_{\mathbf{K},\lambda} t)}\sum_\mathbf{G}\tilde{\bm{\beta}}_{\mathbf{K},\lambda}(\mathbf{G})e^{i {\bf G\cdot r}}\label{eq:bf}
\end{eqnarray}
where the Fourier coefficients $\tilde{\bm{\varepsilon}}_{\mathbf{K},\lambda}(\mathbf{G})$ and $\tilde{\bm{\beta}}_{\mathbf{K},\lambda}(\mathbf{G})$ are given by:
\begin{eqnarray}
\tilde{\bm{\varepsilon}}_{\mathbf{K},\lambda}(\mathbf{G}) &=& -i(\mathbf{K}-\mathbf{G})\tilde{\alpha}^0_{\mathbf{K},\lambda}(\mathbf{G})-i\omega\tilde{\bm{\alpha}}_{\mathbf{K},\lambda}(\mathbf{G})\\
\tilde{\bm{\beta}}_{\mathbf{K},\lambda}(\mathbf{G}) &=& i \mathbf{G}\times \tilde{\bm{\alpha}}_{\mathbf{K},\lambda}(\mathbf{G})
\end{eqnarray}
The symbol for the Fourier coefficients of the electric field $\tilde{\bm{\varepsilon}}$ should not be confused with the dielectric constant $\epsilon(r)$.

\section{Field quantization inside a linear photonic crystal}
\label{sec:quantiz}

Unlike sum frequency generation, parametric down-conversion does
not occur classically and thus is a truly quantum-mechanical
phenomenon. It is thus most natural to discuss the problem in the
language of the quantized electromagnetic field. The procedure for
the quantization of the electro-magnetic field  in a medium
with non-uniform  dielectric differs from that in a uniform dielectric
in that the equations of motion for the
potential~(Eqns.~\ref{eq:pot1}, \ref{eq:pot2}) involve derivatives
of the dielectric function. The two natural generalizations
of the Coulomb gauge: $\epsilon(r) \nabla\cdot\mathbf{A}=0$ and
$\nabla\cdot(\epsilon(r)\mathbf{A})=0$  lead to  different quantization procedures. 
Here we
summarize  the results of Caticha and Caticha~\cite{caticha} who
used the gauge $\epsilon(r) \nabla\cdot\mathbf{A}=0$ to quantize the electro-magnetic field in a medium with periodic dielectric. They showed
that  the Hamiltonian is diagonal in the Bloch-wave basis and the
creation operators for the field satisfy suitably modified
commutation relations. The  quantized field operator is given by:
\begin{equation}
\hat{A}^\mu({\bf r},t)=\sum_{\lambda}\int \frac{\mathrm{d}^3K}{(2\pi)^3} \Big[\hat{a}(\mathbf{K},\lambda) A^\mu_{\mathbf{K},\lambda}({\bf r},t) + h.c.\Big]
\label{eq:ahat}
\end{equation}
Where $A^\mu_{\mathbf{K},\lambda}({\bf r},t) $ is given by Eq.~\ref{eq:amu}. The Hamiltonian can be expressed in the usual form:
\[
\hat{H}=\sum_{\lambda}\int \frac{\mathrm{d}^3K}{(2\pi)^3} \hbar \omega_{\mathbf{K},\lambda} \hat{a}^\dag(\mathbf{K},\lambda)\hat{a}(\mathbf{K},\lambda)
\]
and the creation and annihilation operators $\hat{a}$ and $\hat{a}^\dag$ satisfy the following commutation relations:
\begin{eqnarray}
\big[\hat{a}(\mathbf{K},\lambda),\hat{a}^\dag(\mathbf{K'},\lambda')\big] &=& (2 \pi)^3\delta^{(3)}(\mathbf{K}-\mathbf{K'})\delta_{\lambda,\lambda'} \\
\big[\hat{a}(\mathbf{K},\lambda),\hat{a}(\mathbf{K'},\lambda')\big]&=&0 \\
\big[\hat{a}^\dag(\mathbf{K},\lambda),\hat{a}^\dag(\mathbf{K'},\lambda')\big]&=&0
\end{eqnarray}

The electric and magnetic field operators derived from Eq.~\ref{eq:ahat} are given by:
\begin{eqnarray}
\hat{\mathbf{E}}({\bf r},t)&=&\sum_{\lambda}\int \frac{\mathrm{d}^3K}{(2 \pi)^3} \Big[\hat{a}(\mathbf{K},\lambda) \mathbf{E}_{\mathbf{K},\lambda}({\bf r},t) + h.c.\Big] \label{eq:ehat}\\
\hat{\mathbf{B}}({\bf r},t)&=&\sum_{\lambda}\int \frac{\mathrm{d}^3K}{(2 \pi)^3} \Big[\hat{a}(\mathbf{K},\lambda) \mathbf{B}_{\mathbf{K},\lambda}({\bf r},t) + h.c.\Big]
\end{eqnarray}
with  $\mathbf{E}_{\mathbf{K},\lambda}({\bf r},t)$  and $\mathbf{B}_{\mathbf{K},\lambda}({\bf r},t) $ given by Eqns.~\ref{eq:ef} and~\ref{eq:bf}.
We will now use these results to derive an expression for the quantum interaction Hamiltonian and the phase-matching function.

\section{Non-linear photonic crystal Interaction Hamiltonian and phase-matching}
\label{sec:pm}

To derive the $\chi^{(2)}$  interaction Hamiltonian for  the
quantized electromagnetic field,  we proceed as in the case of
non-linear optical crystals with uniform dielectric and
$\chi^{(2)}$~\cite{pdcth}.  Starting with the expression for  the
classical interaction Hamiltonian:
\[
\epsilon_0  \int \mathrm{d}V \frac{1}{2} \chi^{(2)}_{ijk}(r)E^i(r)E^j(r)E^k(r),
\]
(where  $\chi^{(2)}$ is the second order susceptibility tensor),
 inserting the expression for the quantized electric Bloch-field $\hat{\mathbf{E}}(r,0)$~(Eq. \ref{eq:ehat}),
 making the rotating wave approximation and labelling the three interacting Bloch modes by $p,1,2$, we obtain the following quantum interaction Hamiltonian:
\begin{eqnarray}
\hat{H}_{\mathrm{int}} &=& \sum_{\lambda_{p,1,2}}  \int \mathrm{d}^3K_{p,1,2}\ \mathrm{d}V \nonumber \\
&& \epsilon_0 \chi^{(2)}_{ijk}(r)E^{*i}_p(r)E^j_1(r)E^k_2(r)
\hat{a}_{\mathrm{p}} \hat{a}_{\mathrm{1}}^{\dag}
\hat{a}_{\mathrm{2}}^{\dag}
 +\mathrm{h.c.}\label{eq:intham}
\end{eqnarray}
Where $E_{p,1,2}$ is short for $E_{\mathbf{K}_{p,1,2},\lambda_{p,1,2}}(r,0)$ and
$\hat{a}_{p,1,2}$ is short for $\hat{a}(\mathbf{K}_{p,1,2},\lambda_{p,1,2})$.
 The interaction can be seen to mediate two basic processes: one in which a $p$ photon down-converts  into photons 1 and 2 and (h.c.)  in which photons 1 and 2 up-convert to  photon $p$.

To calculate the time evolution of the field under this interaction Hamiltonian, we switch to the interaction picture, and evaluate the first term in the Dyson series expansion for the time evolution of an initial state $\ket{\psi_0}$:
\begin{widetext}
\[
\ket{\psi(t)}=\Bigg(1+\sum_{\lambda_{p,1,2}}\int \frac{\mathrm{d}^3K_{p,1,2}}{(2 \pi)^3} \delta(\omega_p-\omega_1-\omega_2)  \ \epsilon_0 \int \mathrm{d}V \chi^{(2)}_{ijk}(r)E^{*i}_p(r)E^j_1(r)E^k_2(r) \hat{a}_{\mathrm{p}} \hat{a}_{\mathrm{1}}^{\dag}
\hat{a}_{\mathrm{2}}^{\dag}+\mathrm{h.c.} \big)\Bigg)\ket{\psi_0}
\]
\end{widetext}
For given states $p,1,2$, the amplitude for parametric down-conversion into modes 1 and 2 is proportional to the phase-matching function $\Phi(p,1,2)$, given by:
\begin{equation}
\Phi(p,1,2)=\epsilon_0 \int \mathrm{d}V \chi^{(2)}_{ijk}E^{*i}_p(r)E^j_1(r)E^k_2(r)
\label{eq:pmspatial}
\end{equation}

It is instructive to substitute the Fourier expansion of $E_p(r),E_1(r),E_2(r)$~(Eq.~\ref{eq:ef}) and the Fourier expansion of $\chi^{(2)}_{ijk}(r)$ (with Fourier coefficients denoted by $\tilde{\chi}^{(2)}_{ijk}(\mathbf{G})$) into the phase-matching function (Eq.~\ref{eq:pmspatial}) to obtain:
\begin{widetext}
\begin{equation}
\Phi(p,1,2)=\sum_{\mathbf{G}_\chi,\mathbf{G}_p,\mathbf{G}_1,\mathbf{G}_2}  \epsilon_0 \tilde{\chi}^{(2)}_{ijk}(\mathbf{G}_\chi)\tilde{\varepsilon}^{*i}_p(\mathbf{G}_p)\tilde{\varepsilon}^{j}_1(\mathbf{G}_1)\tilde{\varepsilon}^{k}_2(\mathbf{G}_2) \delta^{(3)}(\mathbf{K}_p-\mathbf{K}_1-\mathbf{K}_2+\mathbf{G}_\chi+\mathbf{G}_1+\mathbf{G}_2-\mathbf{G}_p)  \label{eq:pm}
\end{equation}
\end{widetext}
which makes the conservation of Bloch quasi-momentum manifest.

\begin{table}

\begin{tabular*}{86 mm}{c|c}
\multicolumn{2}{c}{\bfseries Phase-matching dictionary} \\
\multicolumn{2}{c}{\hfill} \\
Natural non-linear crystals & Non-linear photonic crystals \\ \hline
Plane waves  & Bloch waves \\
$\mathbf{E}_\lambda (\mathbf{r})=\mathrm{e}^{i \mathbf{k} \cdot \mathbf{r}}$ &
$\mathbf{E}_{\mathbf{K},\lambda} (\mathbf{r})=\mathrm{e}^{i \mathbf{K} \cdot \mathbf{r}} \sum\limits_{\mathbf{G}} \tilde{\varepsilon}_{\mathbf{K},\lambda}(\mathbf{G}) \mathrm{e}^{i \mathbf{G} \cdot \mathbf{r}}$ \\
\\
Momentum conservation & Quasi-momentum conservation \\
$\mathbf{k_1}+\mathbf{k_2}=\mathbf{k_p}$ & $\mathbf{K_1}+\mathbf{K_2}=\mathbf{K_p}+\mathbf{G}$ \\
\\
Natural dispersion \& & Artificial dispersion \& \\
birefringence & form birefringence \\
$\mathbf{k}(\omega,\mathbf{\hat{k}})= n(\omega,\mathbf{\hat{k}})/c$ & $\mathbf{k}(\omega,\mathbf{\hat{k}})=\mathbf{K}(\omega,\mathbf{\hat{k}})$ \\
\\
Tensor properties $\chi^{(2)}$ &  Tensor properties $\chi^{(2)}$ and \\
determine amplitude & Fourier coefficients \\
& determine amplitude \\
\end{tabular*}
\caption{Phase-matching dictionary that shows  how   the main ideas of phase-matching in non-linear optical crystals translate to  non-linear photonic crystals.}
\label{tab:dict}
\end{table}


Equations~\ref{eq:pmspatial} and \ref{eq:pm} are the main result of this section. To first order in $\chi^{(2)}$ a pump photon will down-convert into a superposition of all Bloch-wave pairs that  satisfy conservation of energy ($\omega_1+\omega_2=\omega_p$) and of Bloch quasi-momentum, with an amplitude $\Phi(p,1,2)$ given by  Eqn.~\ref{eq:pmspatial} or Eqn.~\ref{eq:pm}.  An important difference with the phenomenon of up-conversion is the fact that down-conversion involves all modes that phase-match, whereas up-conversion is more constrained:   given two photons to up-convert, there is typically only one mode they can up-convert to.  A parallel between phase-matching in non-linear crystals with uniform dielectric and phase-matching in photonic crystals is drawn in Table~\ref{tab:dict} which contains a dictionary for the main concepts.

The vector equation that expresses the conservation of Bloch quasi-momentum:
\begin{equation}
\mathbf{K}_1+\mathbf{K}_2= \mathbf{K}_p+\mathbf{G}_\chi + \mathbf{G}_1 + \mathbf{G}_2 - \mathbf{G}_p
\label{eq:phase-matching}
\end{equation}
shall be referred to as the phase-matching equation. For a given $\mathbf{K}_p$, solving the equation corresponds to finding the intersection between the dispersion surfaces
$\vert \mathbf{K}_1(\hat{k}_1,\omega_1)\vert $ and $\vert \mathbf{K}_2(\hat{k}_2,\omega_2)\vert $  for photons 1 and 2, centred  on the origin and on  $\vert \mathbf{K}_p(\hat{k}_p,\omega_1+\omega_2)\vert$ where $\hat{k}$ represents a unit vector that points in the same direction as $\mathbf{K}$.
To solve the equation it is therefore necessary to compute the dispersion surfaces of photons with frequencies $\omega_1, \omega_2$ and $\omega_p$.

For photons that satisfy the phase-matching equation, the amplitude of the process is  proportional to the overlap of the waves in the non-linear medium. As we shall see in later sections, this is most easily  calculated  using Eq.~\ref{eq:pmspatial} in cases where the light field has standing wave character   and  using Eq.~\ref{eq:pm} when it has the character of a propagating wave. It is therefore useful to keep both these expressions in mind when solving the phase-matching problem in a given structure.

In deriving  Eq.~\ref{eq:pm}, the photonic crystal (interaction) volume was taken to be infinitely large. This will of course  not be the case for real photonic crystals. For finite-dimensional crystals the delta functions embodying the conservation of momentum become sinc functions:
\[
\frac{\sin\big[(\mathbf{K}_p-\mathbf{K}_1-\mathbf{K}_2+\mathbf{G}_\chi+\mathbf{G}_1+\mathbf{G}_2-\mathbf{G}_p)_{i} L_i\big]}{(\mathbf{K}_p-\mathbf{K}_1-\mathbf{K}_2+\mathbf{G}_\chi+\mathbf{G}_1+\mathbf{G}_2-\mathbf{G}_p)_{i} L_i}
\]
where $L_i$ is the length of the photonic crystal in the $i$th direction.
This corresponds to a certain amount of allowed momentum mismatch and, as will be seen in section~\ref{sec:fff}, can have significant consequences.

In one and two dimensional photonic crystals, there are two main
mechanisms by which the dispersion surfaces are modified by the
presence of the crystal: form birefringence~\cite{formbi} and
geometric dispersion. Although the two effects are not entirely
independent of each other, their origin is physically distinct.
Form birefringence is the difference in the dispersion surfaces of
Bloch waves that have different polarizations and arises from the
different boundary conditions at the interfaces in the photonic
crystal. In the long wavelength limit ($\lambda\gg\Lambda$), this type of dispersion is
the dominating one. The geometric dispersion is induced by the
presence of the periodic potential and appears as the only type of
dispersion for waves propagating in directions for which the
boundary conditions do not break the symmetry between
polarizations. In three dimensional structures the problem is more
complicated since the de-coupling between direction and
polarization does not readily occur.

The derivations so far apply in all dimensions, however from here on we will restrict our attention to one dimensional (Bragg) structures and discuss the analytical solution of the down-conversion phase-matching problem in detail with a view to generating entangled photon pairs. The problem in the two dimensional case is similar, however calculation of  dispersion relations in two-dimensional structures has so far only been approached numerically. For a discussion of phase-matching in two-dimensional photonic crystals see refs~\cite{earlyth}. Ref~\cite{berger} discusses the problem in a structure having a two-dimensional periodic  $\chi^{(2)}$, but uniform linear dielectric.

\section{Bloch waves in (one dimensional) Bragg structures}

\label{sec:yariv}

\begin{figure}
\includegraphics[width=\columnwidth]{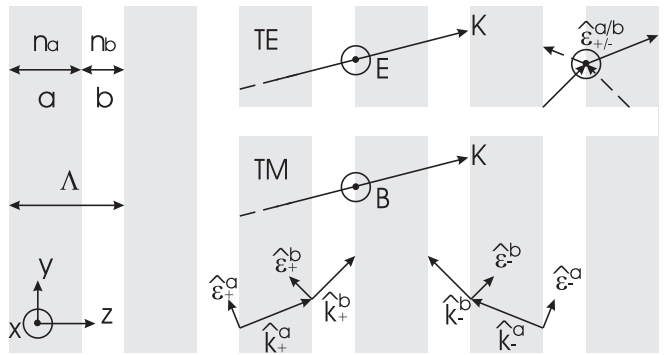}
\caption{Illustration of a one dimensional photonic crystal composed of alternating layers of materials $a$ and $b$ of thickness $a$ and $b$ having refractive indices $n_a$ and $n_b$. The structure is periodic with period $\Lambda=a+b$. The axis of symmetry is taken to be the $z$ axis. There are two types of propagating polarization eigen-modes: Transverse Electric (TE) and Transverse Magnetic (TM). TE(TM) waves have the electric(magnetic) field vector lying in a plane parallel to the interfaces between the materials. } \label{fig:pcsco}
\label{fig:permed}
\end{figure}
In this section we summarize the results of Yariv and
Yeh~\cite{yariv}, describing the Bloch-wave solutions that
propagate in one-dimensional photonic crystals (Bragg structures).
We extend their results by obtaining expressions for the
Bloch-wave Fourier coefficients and discuss the form of the
Bloch-waves in detail, in order to gain intuition that will be
needed in  the discussion of section~\ref{sec:fff}.

 Figure~\ref{fig:permed} shows the basic structure of a one dimensional photonic crystal consisting of a series of alternating layers of materials $a$ and $b$. We adopt Cartesian coordinates with the $z$ axis aligned with the axis of symmetry. Inside each layer the field can be expressed as a superposition of forward and backward propagating plane waves, with wave vectors $\mathbf{k}$ having magnitude $\vert\mathbf{k}\vert=\frac{n\omega}{c}$ (where n is the refractive index in the layer) and component $k^{a,b}_z$ along the $z$ axis. Translational symmetry in the plane perpendicular to the z axis implies, through conservation of momentum, that the components of the wave-vectors parallel to the planes ($k_x$ and $k_y$) are equal across the boundaries, thus the $z$ components are given by:
\begin{equation}
k^{a,b}_z=\sqrt{\Big(\frac{n_{a,b} \omega }{c}\Big)^2-k_{||}^2}\label{eq:kp}
\end{equation}
where $k_{||}^2=k_x^2+k_y^2$ and $n_{a,b}$ are the refractive
indices of materials $a$ and $b$. For waves traveling at an angle
to  the $z$ direction,  there is a natural choice for the
polarization basis: Transverse Electric (TE) with the electric
field pointing out of the plane defined by the wave-vector and the
$z$ axis and Transverse Magnetic (TM), with the electric field
lying in the plane (see Figure~\ref{fig:permed}).

We can thus write down the following expressions for the electric field in the n$^{\mathrm{th}}$ layer:
\begin{eqnarray}
\mathbf{E}^{a}_{n,\mathrm{TE(M)}}=e^{i(\omega t-k_y y)}&\Big(& (a^+_n \hat{\mathbf{\varepsilon}}^a_+)_\mathrm{TE(M)} e^{-ik^a_z(z-n\Lambda )}+\nonumber \\
& & (a^-_n \hat{\mathbf{\varepsilon}}^a_-)_\mathrm{TE(M)}e^{ik^a_z(z-n\Lambda)}\Big)\label{eq:ggg1} \\
\mathbf{E}^b_{n,\mathrm{TE(M)}}=e^{i(\omega t-k_y y)}&\Big(& (b^+_n \hat{\mathbf{\varepsilon}}^b_+)_\mathrm{TE(M)} e^{-ik^b_z(z-n\Lambda )}+\nonumber \\
& & (b^-_n \hat{\mathbf{\varepsilon}}^b_-)_\mathrm{TE(M)}e^{ik^b_z(z-n\Lambda)}\Big). \label{eq:ggg2}
\end{eqnarray}
where $(\hat{\mathbf{\varepsilon}}^{a,b}_{\pm})_{\mathrm{TE(M)}}$ are polarization unit vectors defined in Figure~\ref{fig:permed} and we have chosen $y$ as the off-axis direction for convenience. Matching  the fields at the boundaries we obtain relations between the amplitudes in neighbouring slabs, which can be further reduced to a relation between coefficients in the same material in neighbouring cells:
\[
\left(\begin{array}{c}a^+_{n-1}\\a^-_{n-1}\end{array}\right)=\left(\begin{array}{cc} A & B\\
C&D
\end{array}\right)\left(\begin{array}{c}a^+_{n}\\ a^-_{n}\end{array}\right)
\]
 The coefficients for TE and TM waves are~\cite{yariv}:
 \begin{eqnarray*}
A_{\mathrm{TE}}&=&e^{i k^a_z a}\Big[\cos(k^b_z b)+\frac{i}{2}\Big(\frac{k^b_z}{k^a_z}+\frac{k^a_z}{k^b_z}\Big) \sin(k^b_z b) \Big]\\
B_{\mathrm{TE}}&=&e^{-i k^a_z a}\Big[\frac{i}{2}\Big(\frac{k^b_z}{k^a_z}-\frac{k^a_z}{k^b_z}\Big) \sin(k^b_z b)\Big]\\
A_{\mathrm{TM}}&=&e^{i k^a_z a}\Big[\cos(k^b_z b)+\frac{i}{2}\Big(\frac{n_b^2 k^a_z}{n_a^2 k^b_z}+\frac{n_a^2 k^b_z}{n_b^2 k^a_z}\Big) \sin(k^b_z b)\Big]\\
B_{\mathrm{TM}}&=&e^{-i k^a_z a}\Big[\frac{i}{2}\Big(\frac{n_b^2 k^a_z}{n_a^2 k^b_z}-\frac{n_a^2 k^b_z}{n_b^2 k^a_z}\Big) \sin(k^b_z b)\Big]
\end{eqnarray*}
with $C_{\mathrm{TE/TM}}=B_{\mathrm{TE/TM}}^*$, $D_{\mathrm{TE/TM}}=A_{\mathrm{TE/TM}}^*$.
The eigen-modes of propagation can then be obtained by finding the eigen-vectors  and eigen-values of the transfer matrix.
The right-moving eigen-value is $e^{i K_z \Lambda}$ with:
\begin{equation}
\mathrm{K}_z^{\mathrm{TE/TM}}(k_{||},\omega)=\frac{1}{\Lambda}\cos^{-1}\Big[A_{\mathrm{TE/TM}}+D_{\mathrm{TE/TM}}\Big]
\label{eq:K}
\end{equation}
where $\mathrm{K}_z$ is the $z$ component of $\mathbf{K}=(K_x=k_x, K_y=ky, K_z)$ and the corresponding eigen-vector is:
\begin{eqnarray}
a_n^{\mathrm{TE/TM}}&=&e^{-inK_z^{\mathrm{TE/TM}}\Lambda} B_{\mathrm{TE/TM}}\label{eq:g1}\\
b_n^\mathrm{TE/TM}&=&e^{-inK_z^{\mathrm{TE/TM}}\Lambda} (e^{-iK_z^{\mathrm{TE/TM}}\Lambda}-A_{\mathrm{TE/TM}})\label{eq:g2}
\end{eqnarray}
the $b^\pm_n$ coefficients  are related to the  $a^\pm_n$s via $(b^+_n,b^-_n)=\mathbf{M}(a^+_n,a^-_n)$ where $\mathbf{M}$ is given by:
\begin{equation}
\mathbf{M}_{\mathrm{TE}}=\left(\begin{array}{cc} \frac{(k^a_z+k^b_z)}{2k^b_z}e^{ia(k^a_z-k^b_z)} & \frac{(k^b_z-k^a_z)}{2k^b_z}e^{-ia(k^a_z+k^b_z)} \\
\frac{(k^b_z-k^a_z)}{2k^b_z}e^{ia(k^a_z+k^b_z)} & \frac{(k^b_z+k^a_z)}{2k^b_z}e^{-ia(k^a_z-k^b_z)}
\end{array}\right)\end{equation}
\begin{equation}
\mathbf{M}_{\mathrm{TM}}=\left(\begin{array}{cc} \frac{(n_b^2 k^a_z+n_a^2 k^b_z)}{2n_a n_b k^b_z}e^{ia(k^a_z-k^b_z)} & \frac{(n_b^2 k^a_z-n_a^2k^b_z)}{2n_an_bk^b_z}e^{-ia(k^a_z+k^b_z)} \\
\frac{(n_b^2k^a_z-n_a^2k^b_z)}{2n_an_bk^b_z}e^{ia(k^a_z+k^b_z)} & \frac{(n_a^2k^b_z+n_b^2k^a_z)}{2n_an_bk^b_z}e^{-ia(k^a_z-k^b_z)}
\end{array}\right)\end{equation}

To relate these expressions to the Bloch-wave expressions in
sections~\ref{sec:pcwaves} and~\ref{sec:quantiz} we need to
evaluate the Fourier transform of the expressions for the electric
field~(Eqns.~\ref{eq:ggg1} and~\ref{eq:ggg2}) and compare it to
Eq.~\ref{eq:ef} (with $\mathbf{G}=n\frac{2
\pi}{\Lambda}\hat{\mathbf{z}}$). This is done in
Appendix~\ref{sec:appendix1} where we derive the following
expression for
$\tilde{\bm{\varepsilon}}_{\mathbf{K},\lambda}$:
\begin{eqnarray*}
&& \tilde{\bm{\varepsilon}}_{\mathbf{K},\lambda}\Big(\mathbf{G}=n \frac{2 \pi}{\Lambda}\hat{\mathbf{z}}\Big) \\
&=&\frac{a}{\sqrt{2 \pi}} a^+_0 \frac{\sin[(K_z- k^a_z-n \frac{2 \pi}{\Lambda})\frac{a}{2}]}{(K_z- k^a_z-n \frac{2 \pi}{\Lambda})\frac{a}{2}} e^{-i(K_z- k^a_z-n \frac{2 \pi}{\Lambda})\frac{a}{2}}
\hat{\varepsilon}^a_+ \\
&+& \frac{a}{\sqrt{2 \pi}} a^-_0 \frac{\sin[(K_z+ k^a_z-n \frac{2 \pi}{\Lambda})\frac{a}{2}]}{(K_z+ k^a_z-n \frac{2 \pi}{\Lambda})\frac{a}{2}} e^{-i(K_z+ k^a_z-n \frac{2 \pi}{\Lambda})\frac{a}{2}}
\hat{\varepsilon}^a_- \\
&+&
\frac{b}{\sqrt{2 \pi}} b^+_0 \frac{\sin[(K_z- k^b_z-n \frac{2 \pi}{\Lambda})\frac{b}{2}]}{(K_z- k^b_z-n \frac{2 \pi}{\Lambda})\frac{b}{2}} e^{-i(K_z- k^b_z-n \frac{2 \pi}{\Lambda})(a+\frac{b}{2})}  \hat{\varepsilon}^b_+ \\
&+&
\frac{b}{\sqrt{2 \pi}} b^-_0 \frac{\sin[(K_z+ k^b_z-n \frac{2 \pi}{\Lambda})\frac{b}{2}]}{(K_z+ k^b_z-n \frac{2 \pi}{\Lambda})\frac{b}{2}} e^{-i(K_z+ k^b_z-n \frac{2 \pi}{\Lambda})(a+\frac{b}{2})}  \hat{\varepsilon}^b_- \\
\label{eq:coeffs}
\end{eqnarray*}

\begin{figure}[!t]
\includegraphics[angle=0,width= \columnwidth]{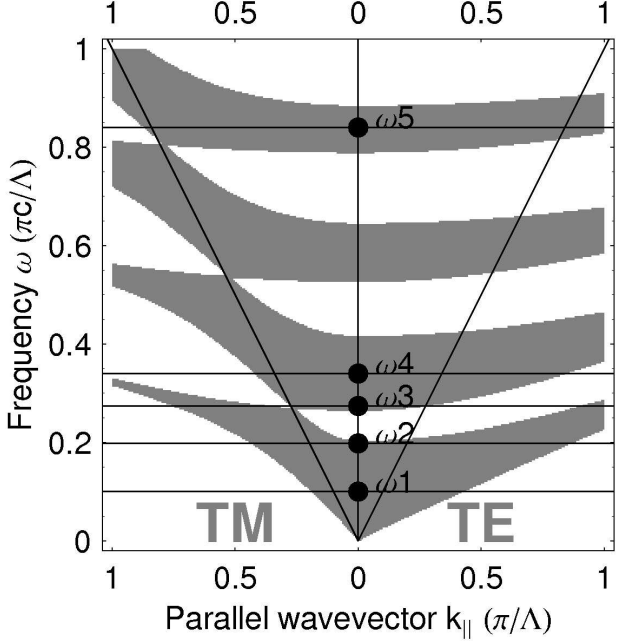}
\caption{Band diagram for a structure of alternating layers of material $a$ with refractive index $n_a=1$ and material b with $n_b=5$. The fill fraction of material $a$: $a/\Lambda$ is $\frac{1}{4}$. For simplicity, natural dispersion and absorption were neglected. Under this simplification, the diagram can be drawn for frequencies and parallel components of $k$ expressed as multiples of $\frac{\pi c}{\Lambda}$ and $\frac{\pi}{\Lambda}$  respectively. The points labelled $\omega_{1\ldots5}$ correspond to the Bloch-wave plots  of Figure~\ref{fig:blwavegallery} and dispersion surface plots  of Figure~\ref{fig:formbi}. } \label{fig:band1}
\end{figure}

 \begin{figure}
\includegraphics[angle=0,width=0.49 \columnwidth]{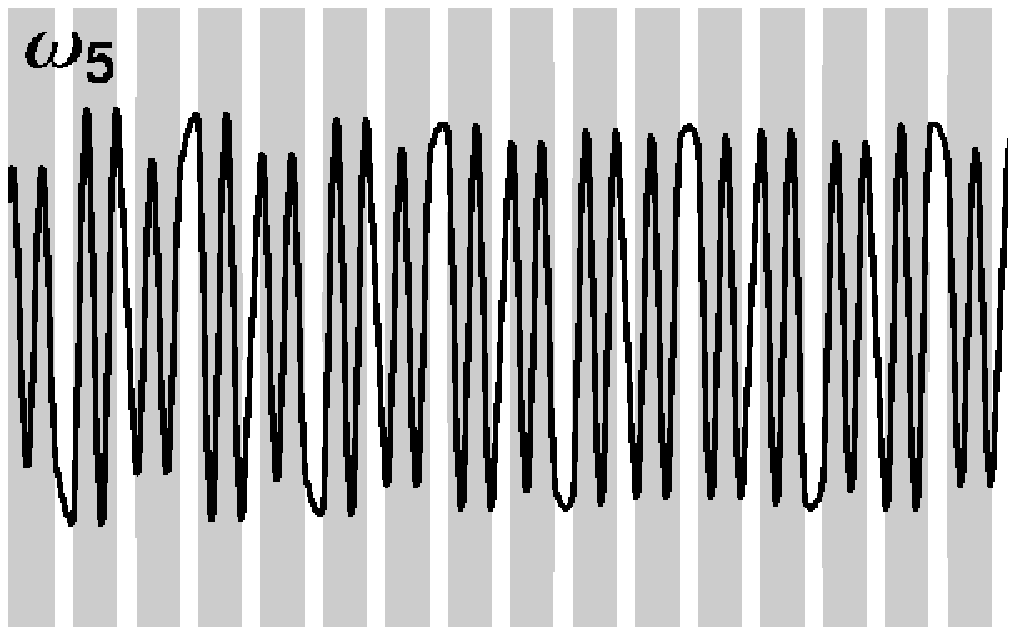} \includegraphics[width=0.49 \columnwidth]{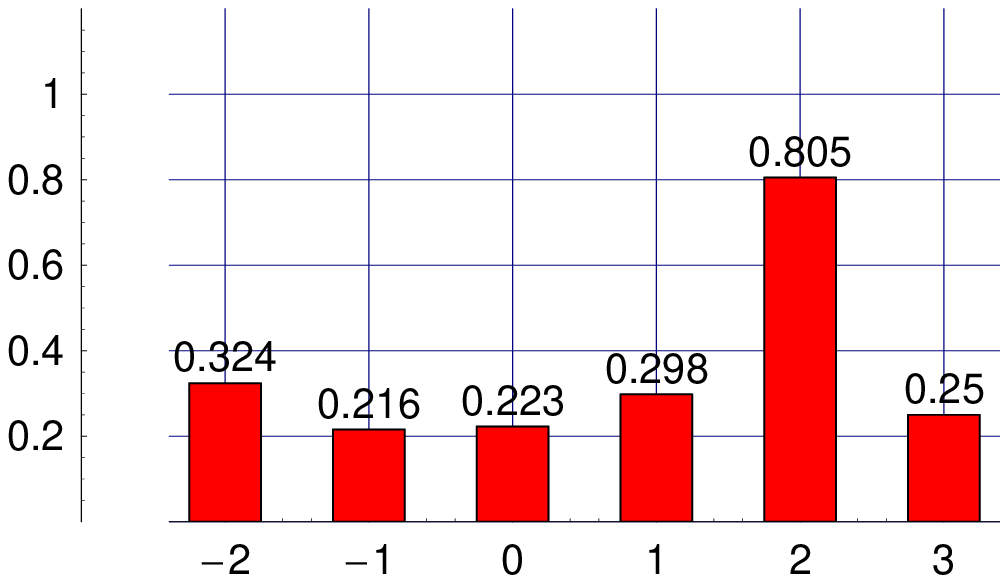}\\
\includegraphics[angle=0,width=0.49 \columnwidth]{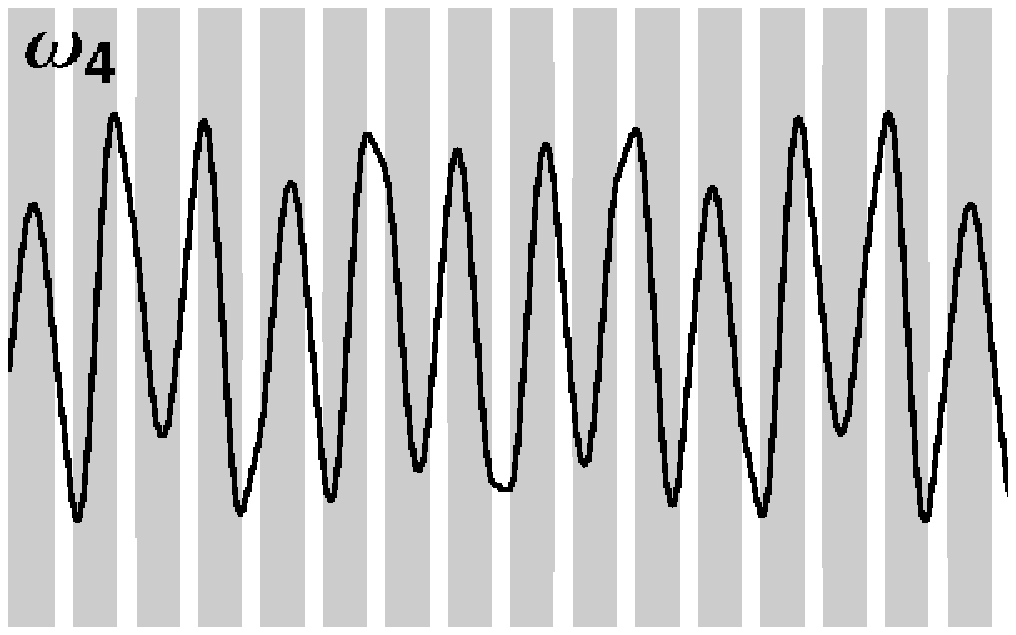} \includegraphics[width=0.49 \columnwidth]{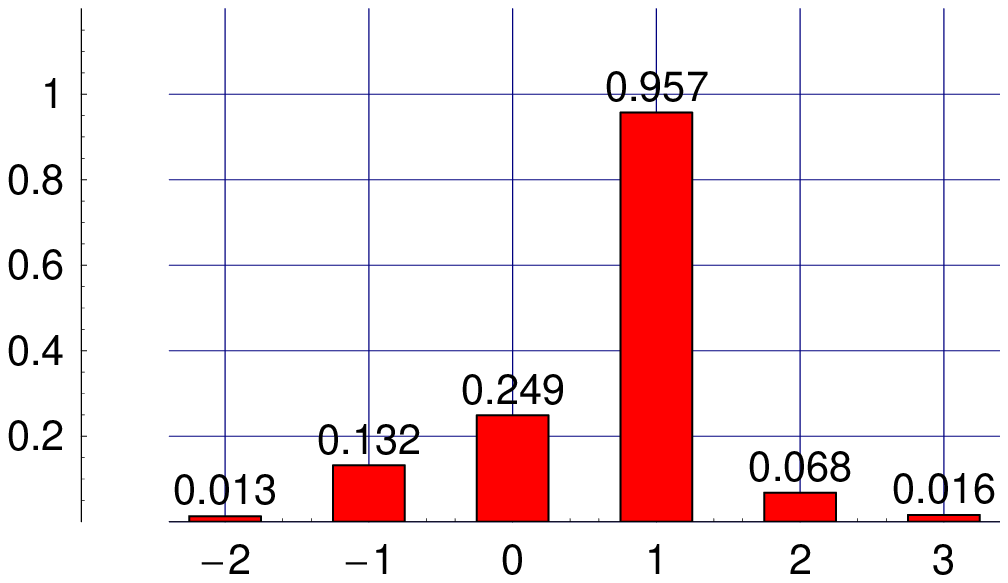}\\
\includegraphics[angle=0,width=0.49 \columnwidth]{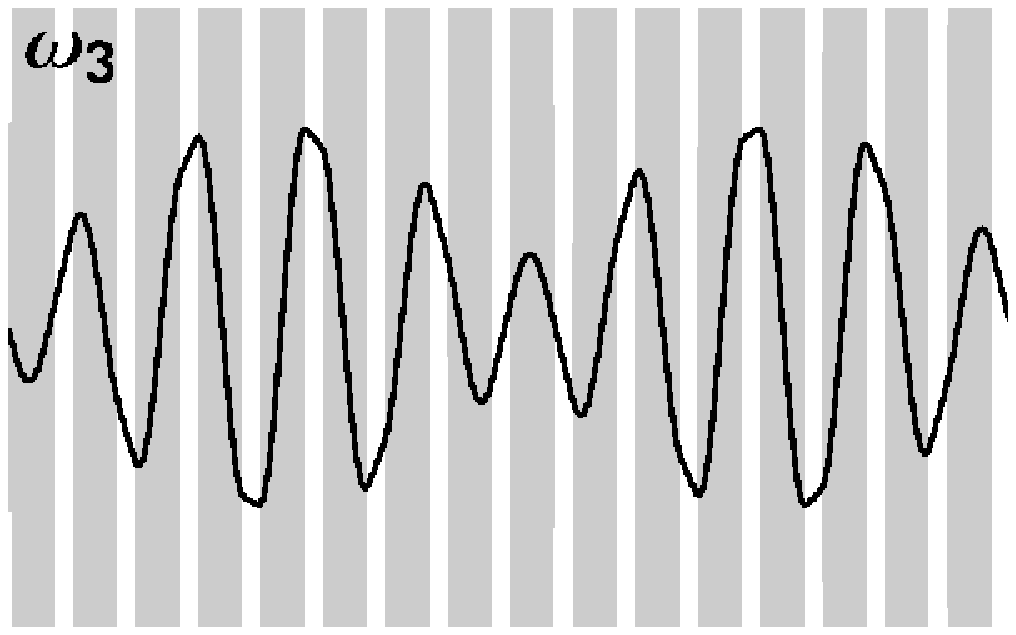} \includegraphics[width=0.49 \columnwidth]{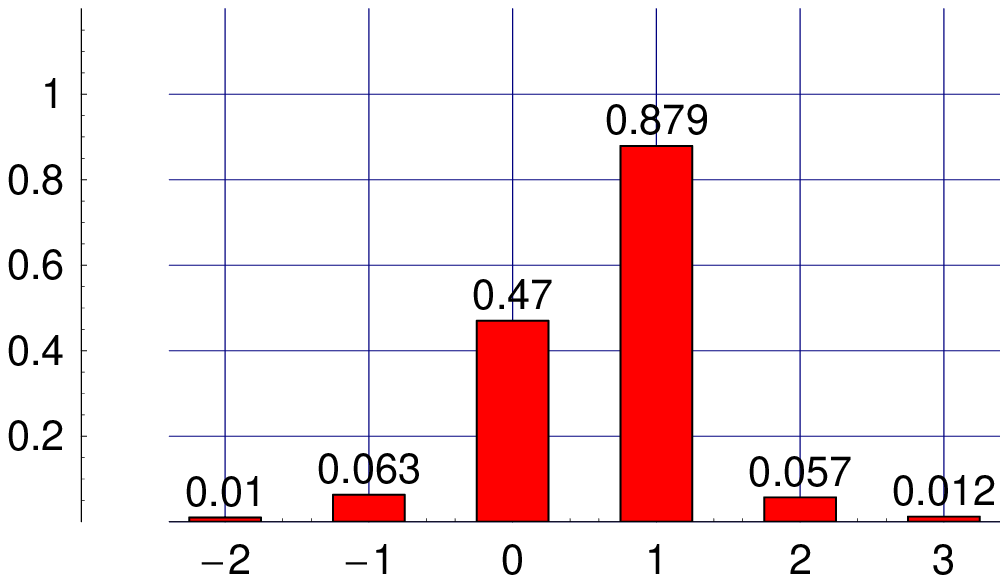}\\
\includegraphics[angle=0,width=0.49 \columnwidth]{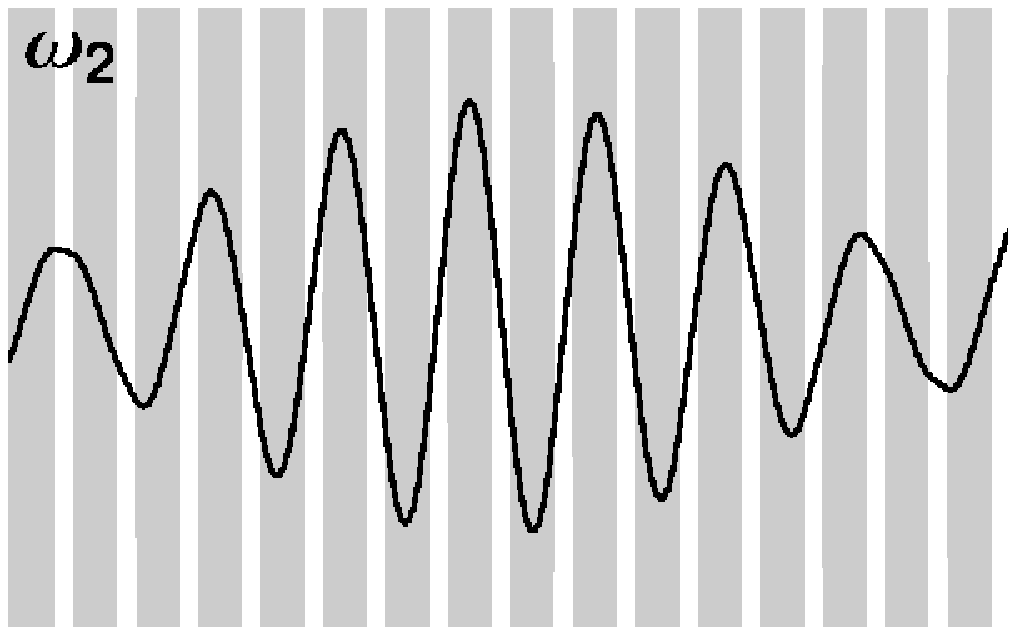} \includegraphics[width=0.49 \columnwidth]{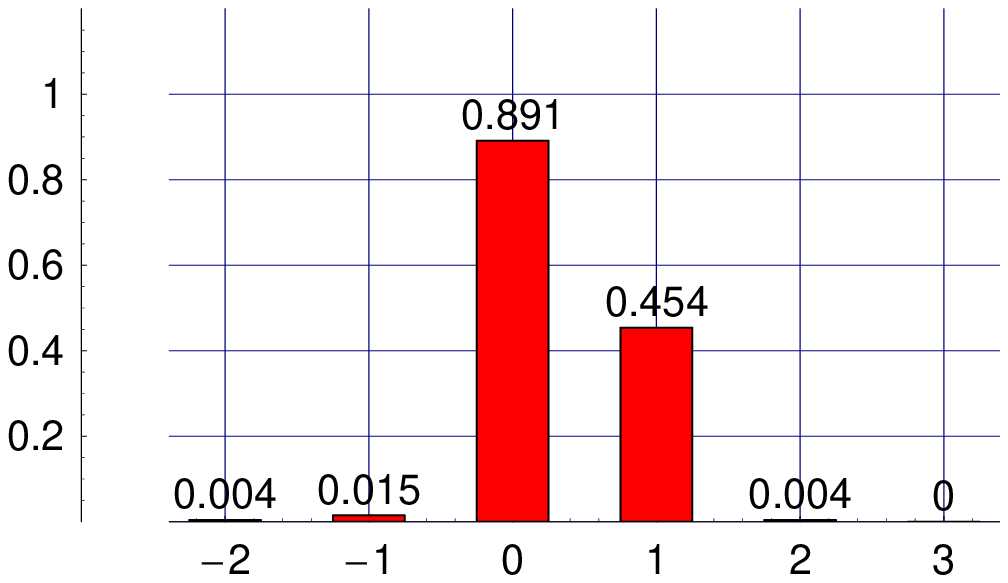}\\
\includegraphics[angle=0,width=0.49 \columnwidth]{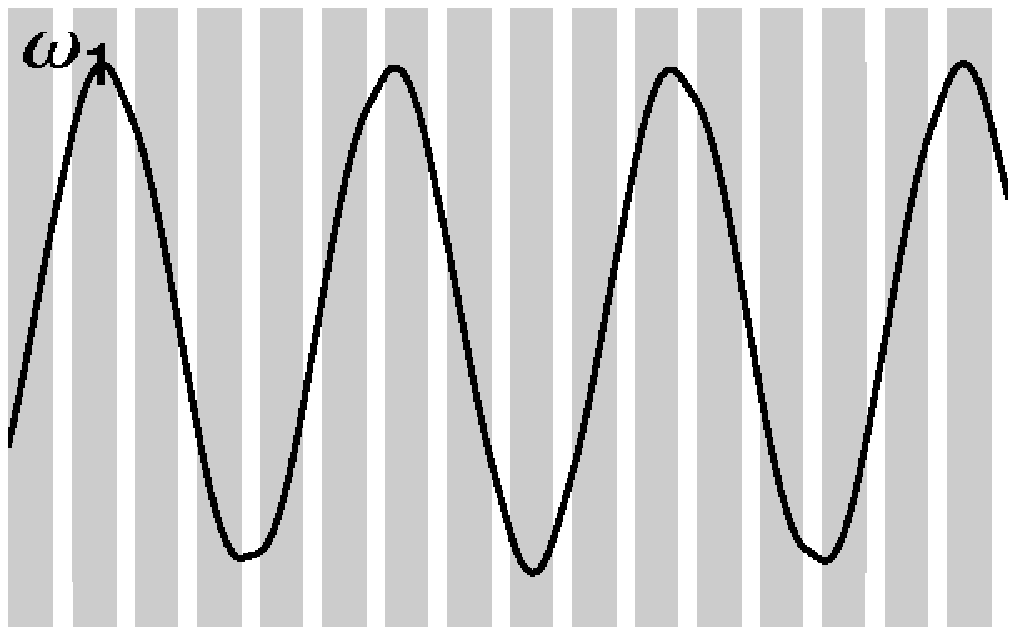} \includegraphics[width=0.49 \columnwidth]{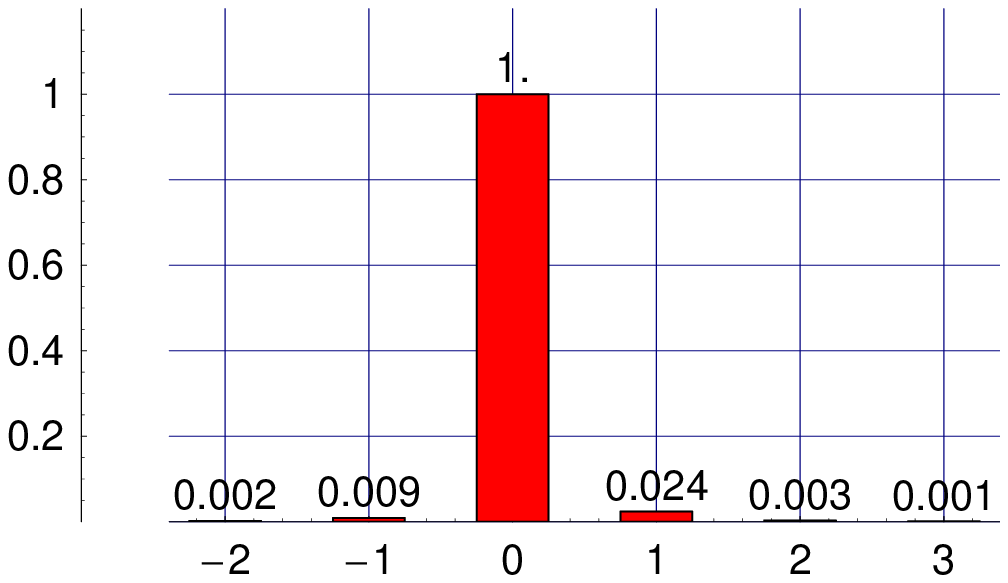}
\caption{({\sc Left:})  Field strength  and ({\sc Right:}) Fourier components of degenerate TE/TM Bloch waves propagating along the axis of symmetry of the illustrative structure with band diagram shown in Figure~\ref{fig:band1}. The wave parameters can be read off Figure~\ref{fig:band1}. Close to the band edge ($\omega_2,\omega_3$) the counter-propagating components grow to make the wave increasingly like a standing wave with the field concentrated in the material of lower or higher refractive index depending on whether the frequency is close to the bottom or the top of a band-gap. As the wavelength gets shorter, the waves deviate more and more from a plane wave even close to the centre of a band ($\omega_5$). } \label{fig:blwavegallery}
\end{figure}

The  expressions for the dispersion relations $\vert \mathbf{\mathrm{K}}_{p,1,2}(\hat{K}_{p,1,2},\omega_{p,1,2})\vert$, the field amplitude $\mathbf{E}_{p,1,2}(r)$, and its  Fourier coefficients $\tilde{\bm{\varepsilon}}_{\mathbf{K}_{p,1,2},\lambda_{p,1,2}}\Big(n \frac{2 \pi}{\Lambda}\Big)$ can be substituted into Equations~\ref{eq:pmspatial} and~\ref{eq:pm} to give
 analytical expressions for  the phase-matched emission from any given (one dimensional) non-linear Bragg structure. The expressions are however not immediately intuitive.
 We therefore proceed to  consider  the results of this section  in various situations to gain some intuition on how the Bloch-wave properties depend on the geometry of the crystal and on the properties of the materials.

Central to an intuitive understanding of  Bloch-waves in a given structure is the band diagram for that structure. Figure~\ref{fig:band1} shows a band diagram for a structure of alternating layers of material $a$ with refractive index $n_a=1$ and material $b$ with refractive index $n_b=5$ (the values were chosen for illustrative purposes). Note that the labels $a,b$ shall also be used to indicate the thicknesses of materials $a,b$.  The fill fraction $a/\Lambda$ of material $a$ is $\frac{1}{4}$. For simplicity, natural material dispersion was neglected; it will be included in the section that follows.  The gray(white)  areas correspond to combinations of $\omega$ and $k_{||}$  that propagate(do not propagate).  A given combination of $\omega$ and $k_{||}$   will propagate if the corresponding $\mathrm{K}_{\mathrm{TE/TM}}(k_{||},\omega)$ is real. This will be the case if $\vert \frac{1}{2}(A+D)\vert\leq1$.  For TE waves $\frac{1}{2}(A+D)$ is given by:
\[
\cos(k^a_z a+k_z^b b)+\Big[1-\frac{1}{2}\Big(\frac{k^b_z}{k^a_z}+\frac{k^a_z}{k^b_z}\Big)\Big] \sin(k^a_z a)\sin(k^b_z b)\\
\]
whereas for TM waves it is given by:
\[
\cos(k^a_z a+k_z^b b)+\Big[1-\frac{1}{2}\Big(\frac{n_b^2 k^a_z}{n_a^2 k^b_z}+\frac{n_a^2 k^b_z}{n_b^2 k^a_z}\Big) \Big]\sin(k^a_z a)  \sin(k^b_z b).
\]
It can be seen that the extent to which $\mathrm{K}_z \Lambda$ deviates from the simple linear relation $\mathrm{K}_z \Lambda = k^a_z a+k_z^b b$  depends in a non-trivial way on the ratios of the refractive indices and on the width of the layers compared to the wavelength inside the medium. The term in the square brackets is independent of the width of the layers and is a measure of the strength of the modification of the dispersion relation (and thus the width of the bands). The trigonometric terms in turn depend  on the layer widths in a simple way and have the primary function of determining the positions of the bands.
It is instructive to consider the band-structure for $k_{||}=0$ of a crystal of fixed optical periodicity $l=n_a a+n_b b$ and varying optical fill fraction $f=n_a a/l$. Measuring the free space wavelength as a fraction of $l$: $\lambda_{\mathrm{free\ sp.}}=x\times l$ we obtain the following expression for $(A+D)$:
\[
\cos\Big(\frac{2\pi}{x}\Big)+\Big[1-\frac{1}{2}\Big(\frac{n_a}{n_b}+\frac{n_b}{n_a})\Big]\sin\Big(\frac{2\pi f}{x}\Big)\sin\Big(\frac{2\pi(1-f)}{x}\Big).
\]
It can be seen that the effects of geometric dispersion disappear for
$f/x=n$ or $(1-f)/x=m$ where $n$ and $m$ are integers. This corresponds to regions
in which the scattered waves interfere destructively in the backward direction
and the dispersion becomes the trivial one: $K\Lambda=k_a a+k_b b$.
In these regions, the Bloch-wave is essentially a plane wave (i.e. it has only one dominant Fourier
component).

 \begin{figure}
\includegraphics[angle=0,width= 0.49 \columnwidth]{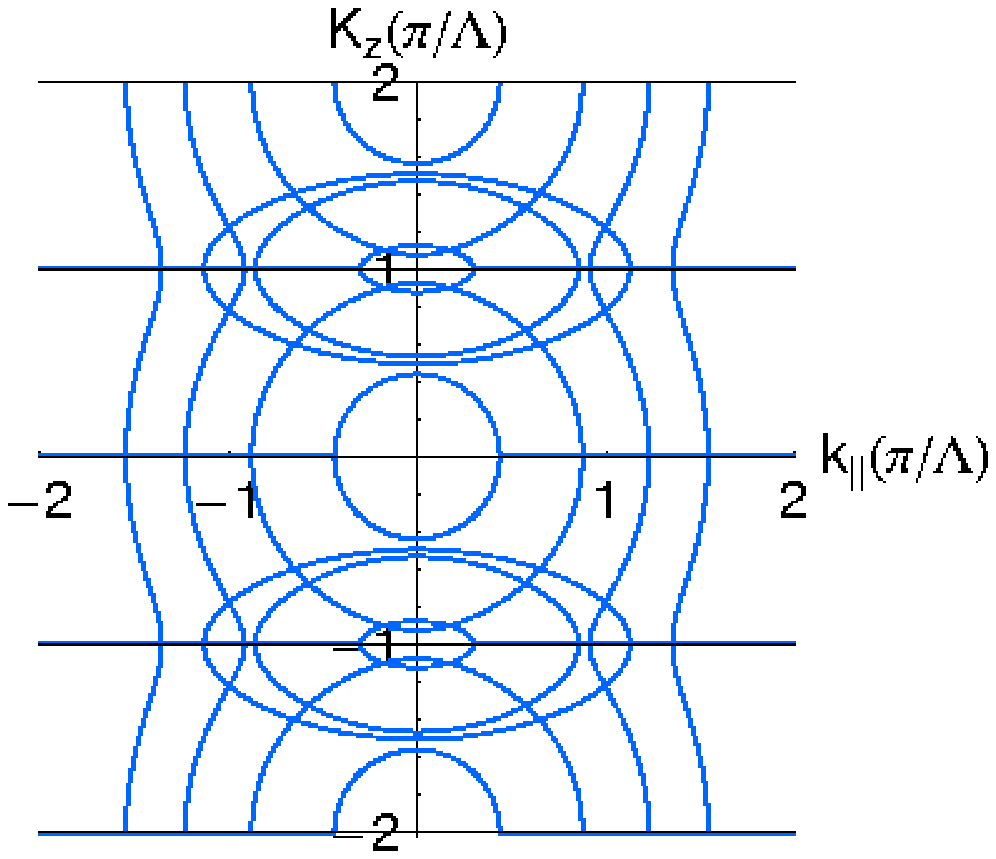} \includegraphics[angle=0,width=0.49 \columnwidth]{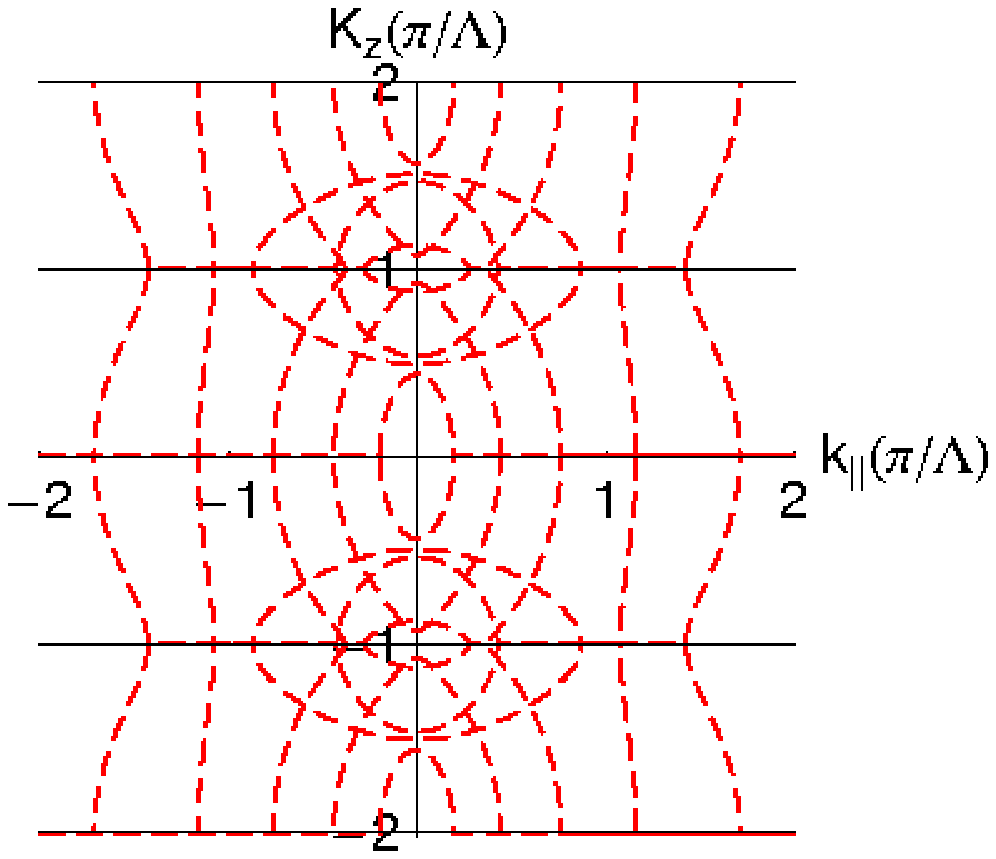}\\
\includegraphics[angle=0,width= 0.49 \columnwidth]{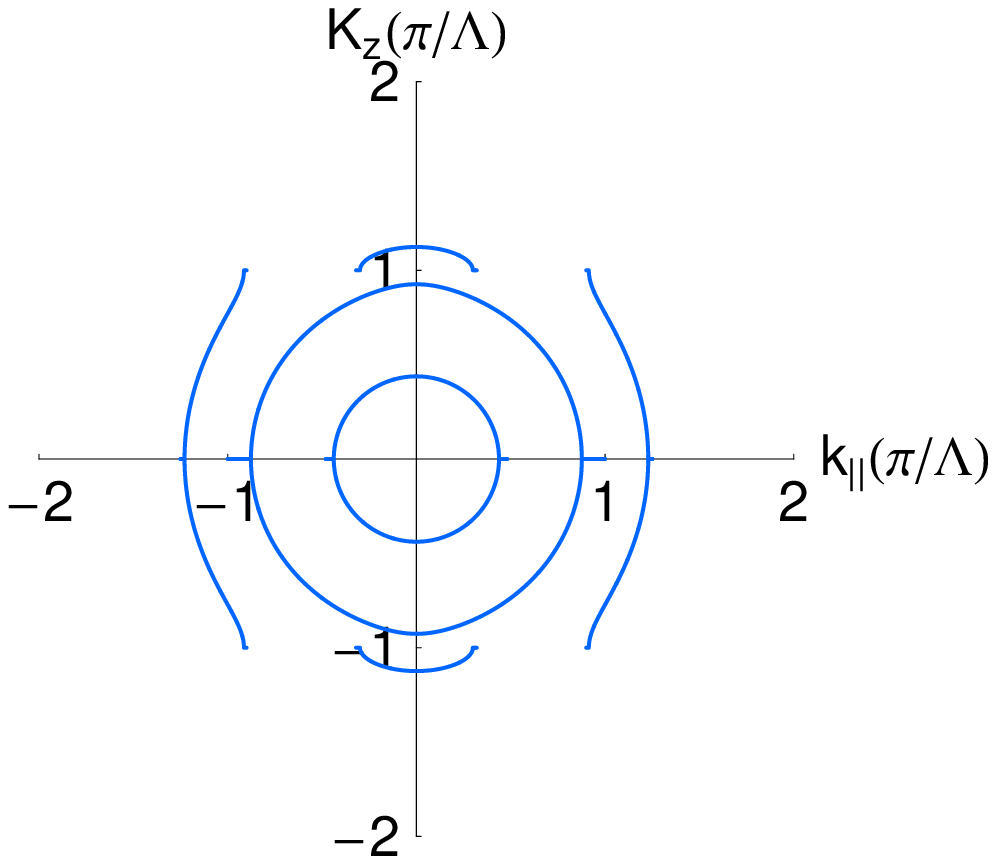} \includegraphics[angle=0,width=0.49 \columnwidth]{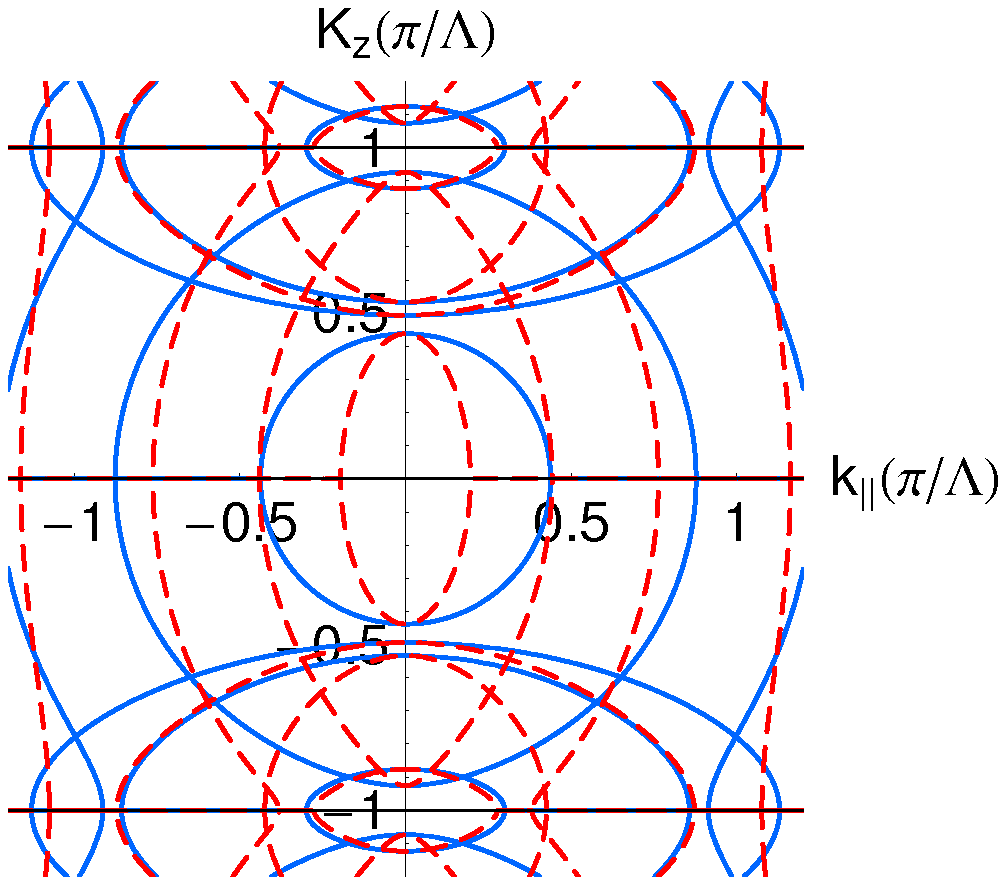}
\caption{Dispersion surfaces for (Top, Left):  TE and (Top, Right): TM  Bloch waves with frequencies $\omega_{1\ldots5}$. (Bottom, Left):  Dispersion surface for the leading TE wave components of waves $\omega_{1\ldots3}$. (Bottom, Right): The dispersion surfaces for TE and TM waves  shown together. In the long wavelength limit the surfaces resemble those of a uniaxial birefringent crystal. As the wavelength is decreased to a length comparable to the periodicity of the  structure, the surfaces increasingly distort.} \label{fig:formbi}
\end{figure}

Figure~\ref{fig:blwavegallery} (Left) shows   the electric field
amplitude and (Right) the Fourier components of Bloch waves that
correspond to the dots in Figure~\ref{fig:band1}. As we approach
the band edges the Bloch waves can be seen to differ  strongly
from plane waves. At the edge of a stop-band  the
$\omega_2$($\omega_3$) waves take the form of standing waves with
the electric field concentrated in the material of higher(lower)
refractive index. The magnitude of $\mathbf{K}$ is the same at
these two frequencies. When estimating the down-conversion
amplitude  it is useful to be able to switch between configuration
and Fourier space depending on the degree of localization of the
Bloch waves. For waves with non-zero $k_{||}$ both the interaction
strength and the position of the bands will change. In particular
the difference between TE and TM waves will become increasingly
marked, as can be seen by inspection of the band structure of
Figure~\ref{fig:band1}. Some intuition on how the magnitude of the
Bloch momentum of propagating waves varies as a function of
$k_{||}$  can be gained by examining Figure~\ref{fig:formbi},
where we show the dispersion surfaces for waves of frequencies
$\omega_{1\ldots4}$. In particular it is apparent that in the long
wavelength limit, the surfaces resemble those for a uniaxial
birefringent crystal. The analogy can be formalized~\cite{yariv}
by expanding the expression for $K(k_{||},\omega)$
(Eq.~\ref{eq:K}) in the limit $\lambda \gg \Lambda$ to obtain
expressions for the corresponding ordinary (TE) and extraordinary (TM)
refractive indices $n_o$ and $n_e$:
\begin{eqnarray}
n_o^2&=&\frac{a}{\Lambda} n_a^2+\frac{b}{\lambda}n_b^2\\
\frac{1}{n_e^2}&=&\frac{a}{\Lambda} \frac{1}{n_a^2}+\frac{b}{\lambda}\frac{1}{n_b^2}
\end{eqnarray}
For phase-matching between waves with different polarization in
the long-wavelength limit, optimization of the birefringence can
provide a method to choose the fill fractions of a structure.

Having gained some intuition on how the properties of Bloch-waves depend on the photonic-crystal geometry and constituent materials we now turn to the problem of calculating the down-converted emission from real structures.

\section{The phase-matching problem in real structures}
\label{sec:fff}

  \begin{figure}
\includegraphics[width=\columnwidth]{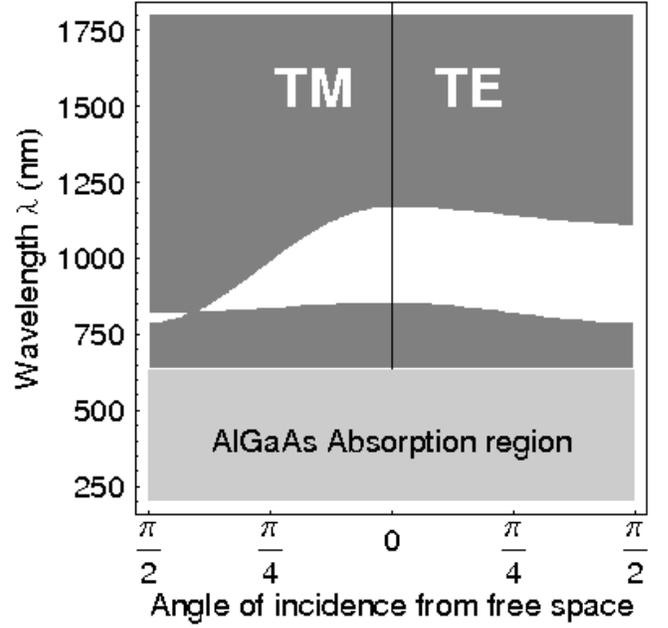}
\caption{Band diagram for the example Al$_{0.4}$Ga$_{0.6}$As/Air structure with periodicity $\Lambda=187.5$nm.
Points in the grey bands correspond to propagating TM (Left) and TE (Right) states. Points in the white bands correspond to frequencies that cannot propagate in the structure. The shaded area below 640nm  represents the region in which Al$_{0.4}$Ga$_{0.6}$As is absorbing.}
\label{fig:band}
\end{figure}
\begin{figure}
\includegraphics[angle=0,width=0.49 \columnwidth]{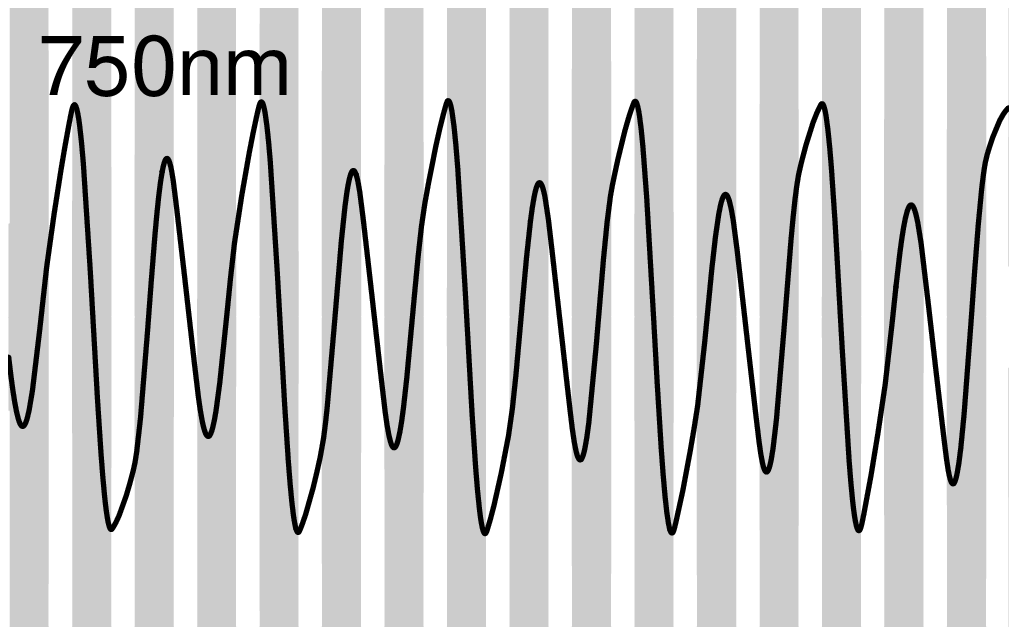} \includegraphics[width=0.49 \columnwidth]{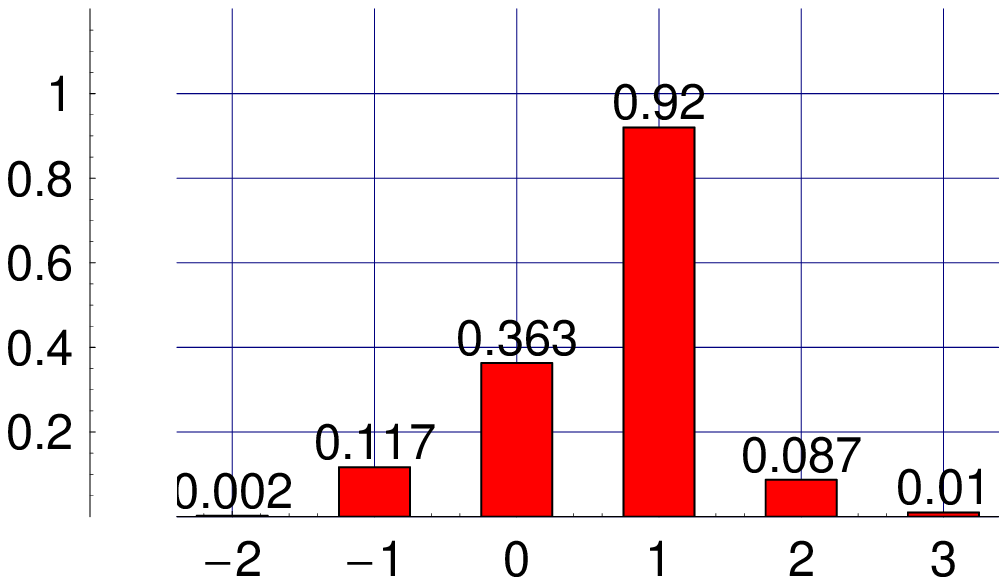}\\
\includegraphics[angle=0,width=0.49 \columnwidth]{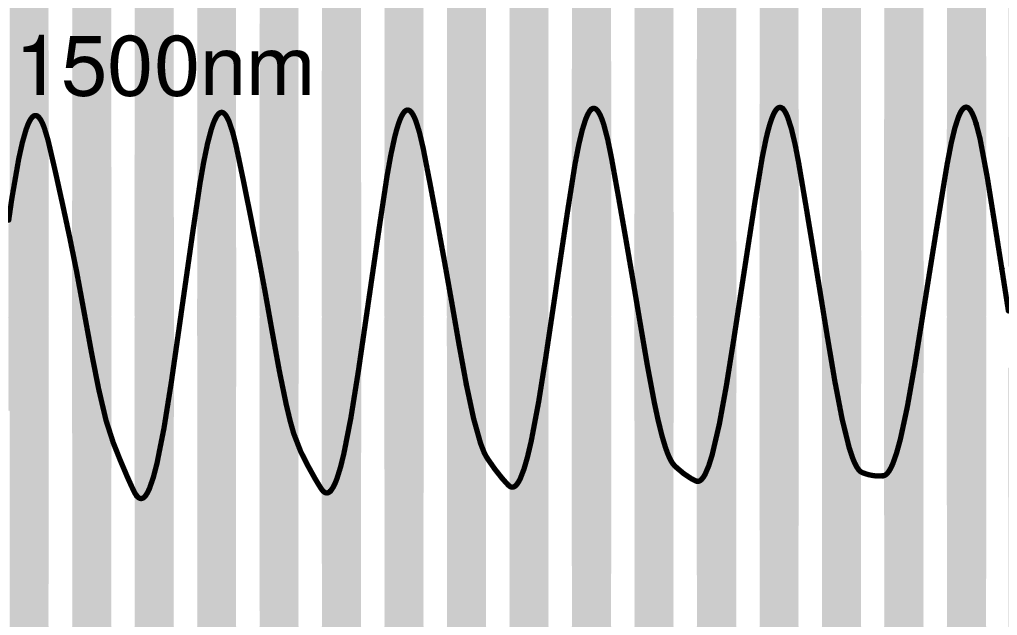} \includegraphics[width=0.49 \columnwidth]{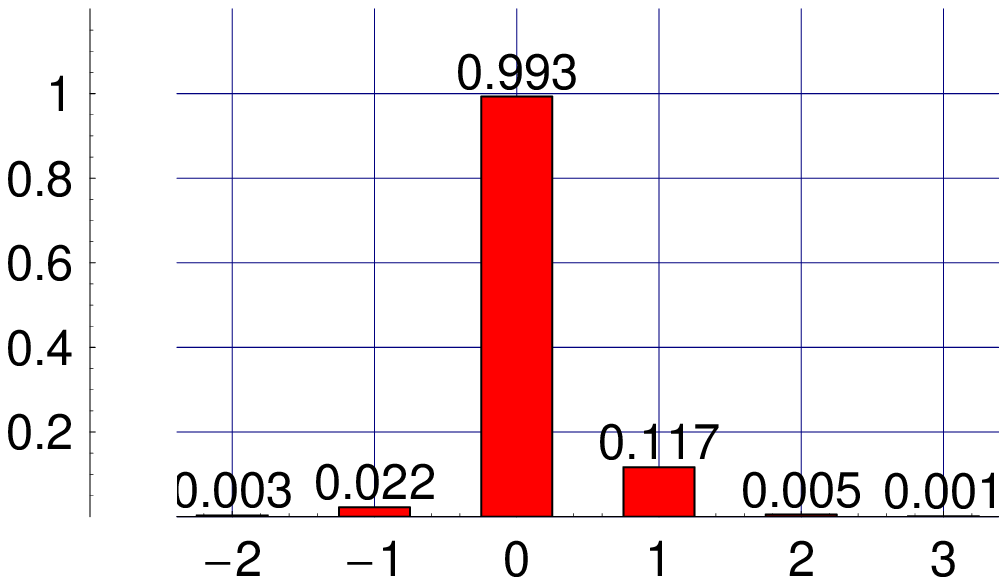}
\caption{({\sc Left}):  Field strength  and ({\sc Right}):
Fourier components of degenerate TE/TM Bloch waves propagating along
the axis of symmetry of the example structure.  } \label{fig:blex}
\end{figure}

The parameter space for the design of a structure that achieves a
particular type of phase-matching is large. Here we give a
procedure to make predictions for the emission. We suggest
iterating the procedure, making changes to the structure guided by
the intuition developed in the previous section, as a means of
optimization.  We take as an example a structure composed of 30
alternating layers of  Al$_{0.4}$Ga$_{0.6}$As (123nm thick) and
air (64.5nm thick) and consider down-conversion of  750nm photons
to  degenerate 1500nm photon pairs. The fraction of Aluminium was
chosen to avoid absorption of the pump photons.  The relative
thickness of the layers was chosen to optimize the birefringence
in the long wavelength limit.  The periodicity was chosen to
ensure both pump and down-converted photons would propagate, the
pump in the second band, the down-converted photons in the first
band. We proceed according to the following recipe:

(i) {\bf Band structure}: We first plot the band diagram (Figure~\ref{fig:band}) of the structure. An intuitive variable in terms of which to plot the band diagram is the angle $\theta$ of propagation of the plane-wave solution inside the material of lower refractive index which is given by: $\sin(\theta)=\frac{k_{||}}{k_z}$. When $\vert \frac{k_{||}}{k_z}\vert > 1 $ total internal reflection occurs at the interfaces between the layers and the Bloch waves do not propagate along the crystal, but rather in a direction perpendicular to the crystal axis.  In the case of the example structure this occurs when $k_{||}$ is equal to the magnitude of the free-space wave-vector $\frac{\omega}{c}$.
  It is of course necessary to include natural absorption of the materials in the band diagram plot since it can  restrict the accessible bands.
   Al$_{0.4}$Ga$_{0.6}$As absorbs at wavelengths below 640nm~\cite{algas}. The natural dispersion of the materials is accounted for  by including the frequency dependence of $n_{a,b}$ (See Ref.~\cite{algas} for AlGaAs) in Eq.~\ref{eq:kp}.

(ii) {\bf Fourier spectrum}:  Having chosen frequencies that propagate (or having varied the periodicity to ensure they propagate), we look at the amplitude of the Fourier components of the Bloch-waves having $k_{||}=0$. Figure~\ref{fig:blex} shows the waves and Fourier components for our example structure. Both have a leading Fourier component. The phase-matched emission will in general not have $k_{||}=0$. However if the combination $\omega,k_{||}=0$ does not lie close to a band edge, then the $k_{||}=0$ spectrum is a good indication of what the $k_{||}\neq0$ spectrum will look like.

\begin{figure}[!t]
\includegraphics[width=\columnwidth]{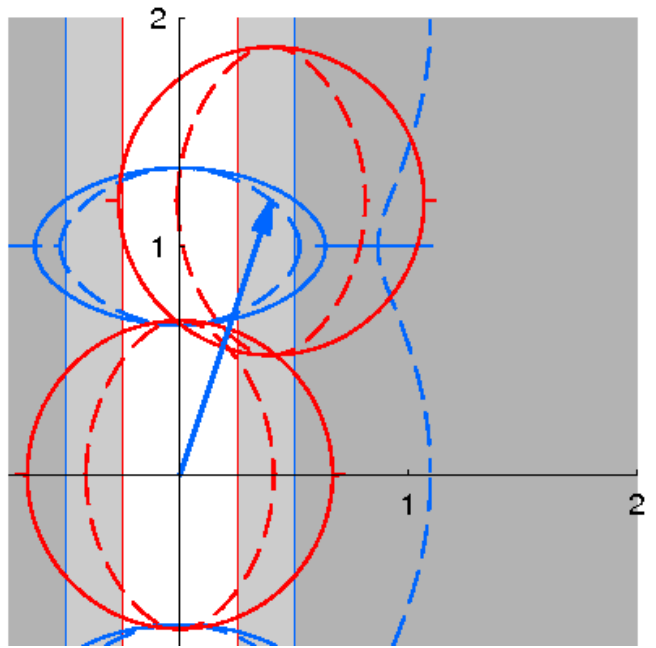}
\caption{Diagram to determine whether down-converted light will
phase-match. The solid(dashed) blue lines represent the dispersion
surfaces for pump ($\lambda=750$nm) TE(TM) photons.  The
solid(dashed) red lines represent the dispersion surface for
down-converted ($\lambda=1500$nm) TE(TM) photons. The pump
dispersion surface is centered at the origin, and the blue arrow
represents the pump Bloch-vector. Two down-converted  photon
dispersion surfaces are drawn, centered on the extremities of the
blue arrow. Intersections between the down-converted photon
surfaces represent Bloch vectors that phase-match according to
Eq~\ref{eq:phase-matching}.  }\label{fig:pm}
\end{figure}

(iii) {\bf Phase-matching of leading terms}: We then look to satisfy the phase-matching equation $\mathbf{K}_p=\mathbf{K}_1+\mathbf{K}_2+\mathbf{G}$ starting with  the leading $\mathbf{G}=n\frac{2\pi}{\Lambda}\hat{\mathbf{z}}$'s from the Fourier analysis above. This is done by drawing a dispersion diagram such as the one shown in Figure~\ref{fig:pm}.
Intersections between the displaced down-converted photon dispersion  surfaces represent Bloch vectors that phase-match according to Equation~\ref{eq:phase-matching}.

The full set of solutions corresponds to the intersection of the full two-dimensional dispersion surfaces.  However the existence of an intersection in the diagrams discussed above is a necessary and sufficient condition for the two surfaces to have an intersection.

\begin{figure}[!t]
\includegraphics[width=0.9\columnwidth]{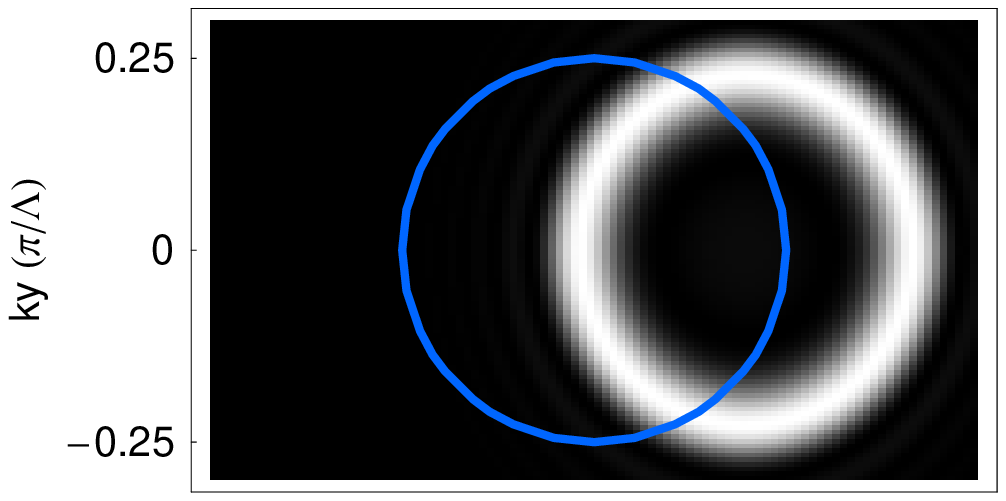}\\
\includegraphics[width=0.9\columnwidth]{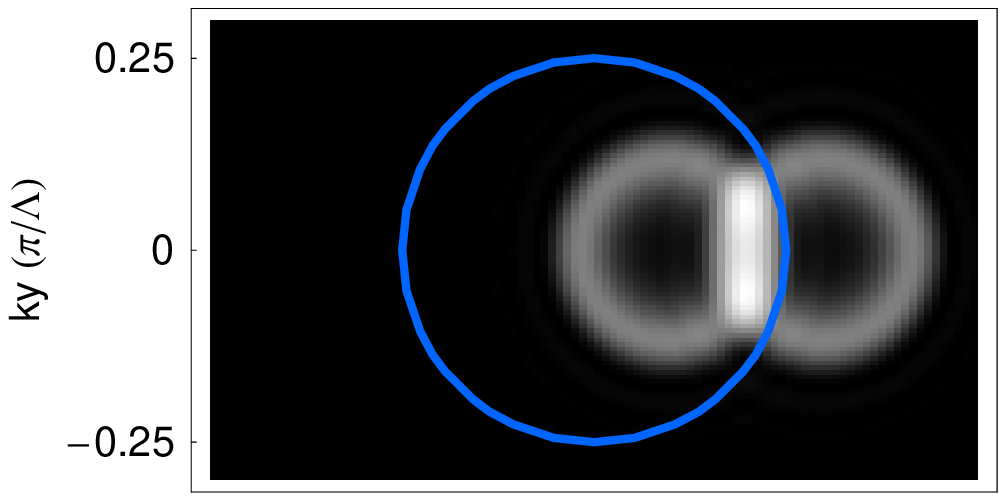}\\
\includegraphics[width=0.9\columnwidth]{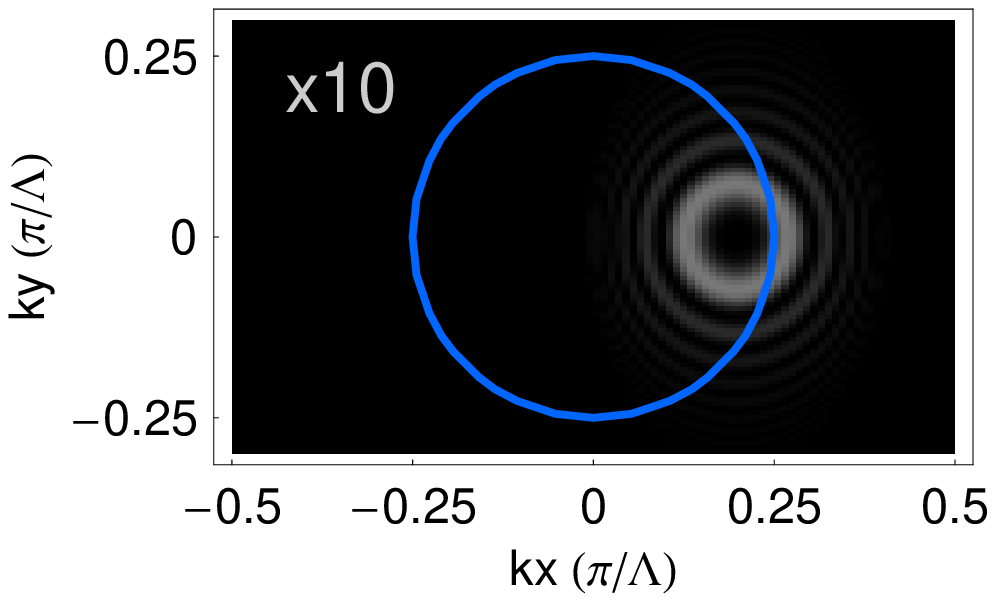}
\caption{Three types of phase-matching in the Al$_{0.4}$Ga$_{0.6}$As/Air structure. (Top) Type-I (TM$\rightarrow$TE+TE), (Middle) Type II (TM$\rightarrow$TE+TM), (Bottom) Type III (TM$\rightarrow$TM+TM). The density plots in $k_{||}$ space represent the emission strength calculated using only the phase-matching  function and do not include contributions from the tensor nature of the $\chi^{(2)}$ non-linearity.  The blue circle represents the values of $k_{||}$ that totally internally reflect between the layers. The Type III phase-matching is weak when compared to types I and II for this set of parameters and so was multiplied by a factor 10 to make it visible.} \label{fig:pmfs}
\end{figure}

(iv) {\bf Plotting the emission}: Having determined that the intersections exist, we look at the detailed direction of the emission. To do this analytically we make the simplifying assumption that the crystal transverse dimensions are infinite ($L_x,L_y\gg\lambda/n_{a,b}$). This is both justified in most experimental situations and makes it possible to produce a manageable analytical prediction of the emission. The transverse part of the phase-matching function then becomes a product of two delta functions that ensure:
\begin{eqnarray*}
k^{(1)}_{x}&=&-k^{(2)}_x \\
k^{(1)}_y &=& k^{(p)}_y-k^{(2)}_y.
\end{eqnarray*}
Substituting these relations into the longitudinal part of the
phase-matching function (Eq.~\ref{eq:pm} with the
$\delta$-function replaced by the appropriate sinc function) to
eliminate either $k^{(1)}_{x,y}$ or $k^{(2)}_{x,y}$, we can then
plot the phase-matching function for each emission. The
elimination procedure is equivalent to calculating a partial trace
over the emissions of the twin photon. It should be noted that the
partial trace calculation  is much simpler than in the case of
birefringent non-linear crystals such as BBO since the optic axis
is by default aligned with the crystal axis. Figure~\ref{fig:pmfs}
shows such a plot for three types of processes: Type I where a
pump photon with TM polarization down-converts to photons 1 and 2
with TE polarization: $p(TM)\rightarrow1(TE)+2(TE)$, Type II:
$p(TM)\rightarrow1(TE)+2(TM)$ and Type III:
$p(TM)\rightarrow1(TM)+2(TM)$. The width of the emission rings
corresponds to the allowed longitudinal momentum mismatch and is
inversely proportional to the crystal length. As can be seen in
Figure~\ref{fig:pmfs}, it is an important parameter for the
crystal size being considered here. For a discussion of the
relation between the crystal thickness and the yield of entangled
photon pairs in BBO, see~\cite{martin}.

(v) {\bf Efficiency of the process}: From the plots of the detailed emission, the efficiency for emission in directions of interest can be estimated. The values of $k_{||}$ for photons 1 and 2 are read off the plot.
 The amplitude for the  phase-matched process can then be calculated  by performing the sum over the Fourier coefficients (Eq.~\ref{eq:pm}), or by numerical integration of the field amplitudes in the non-linear layers  (Eq.\ref{eq:pmspatial}). In the case of the Fourier amplitude calculation it will be necessary to use the coefficients $\tilde{\chi}(\mathbf{G})$ from the Fourier
 expansion of $\chi(r)$ which is:
\begin{eqnarray*}
\chi^{(2)}(r)=\frac{1}{\pi} \sum_n \frac{1}{n} &\Big[& \chi_be^{-i\frac{b\pi n}{\Lambda}}\sin\big(\frac{b\pi}{\Lambda}n\big) \\
&+& \chi_a e^{-i\frac{a\pi n}{\Lambda}}\sin\big(\frac{a\pi}{\Lambda}n\big)\Big]e^{i\frac{2\pi x}{\Lambda}n}
\end{eqnarray*}
The  $n=0$ term is simply the weighted average of the $\chi$s: $\frac{a \chi_a+b\chi_b}{\Lambda}$.

For photons emitted in the directions corresponding to the intersection of the rings in Figure~\ref{fig:pmfs} (Middle) or Figure~\ref{fig:entang} there is  one dominant combination of $\mathbf{G}$s: $\mathbf{G}_\chi=0$, $\mathbf{G}_p=\frac{2 \pi}{\Lambda}\hat{\mathbf{z}}$ and  $\mathbf{G}_1=\mathbf{G}_2=0$. For this combination the Fourier amplitudes are 0.66, 0.90, 0.99 and 0.98 respectively. The combined effect of the Fourier amplitudes leads to a  factor 0.58. 

In addition, the tensor nature of the $\chi^{(2)}$ interaction needs to be included. This is done by taking the contraction of the unit polarization vectors of the relevant Fourier components of $E_p,E_1,E_2$ with the $\chi^{(2)}$ tensor: $\chi^{(2)}_{ijk}(\hat{\varepsilon}^b_p(k^p_x,k^p_y))^i(\hat{\varepsilon}^b_{(1)}(k^{(1)}_x,k^{(1)}_y))^j(\hat{\varepsilon}^b_{(2)}(k^{(2)}_x,k^{(2)}_y))^k$.
The
$\chi^{(2)}$ tensor of AlGaAs has $\bar{4}3m$ point group symmetry
and has only three non-zero coefficients all having magnitude $200 \mathrm{pm/V}$~\cite{chi2}. 
For the conventional (100) surface
orientation, the crystalline axis coincides with the direction of
normal incidence that we have defined to be the $z$ direction. 
For photons emitted in the directions corresponding to the intersection of the rings in Figure~\ref{fig:pmfs} (Middle) or Figure~\ref{fig:entang} this leads to a value of $0.53\times 200\mathrm{pm/V}$.

We are thus in a position to compare the efficiency of the process to that in BBO.  The overall down-conversion efficiency for our example structure and choice of pump parameters is  $(0.53 \times 0.58 \times 200)^2\sim 780$ times that  in a BBO crystal of similar size, where we have assumed that there is no reduction of the $2.2\mathrm{pm/V}$ value of the  BBO $\chi^{(2)}$ coming from the tensor nature of the interaction. 

It should be noted that there will also be down-conversion into directions that phase-match but do not involve leading Fourier amplitudes. These can be treated in exactly the same way, but will have much lower amplitude. 

(vi) {\bf Extraction of polarization-entangled photon pairs}:
\begin{figure}[!t]
\includegraphics[width=0.9\columnwidth]{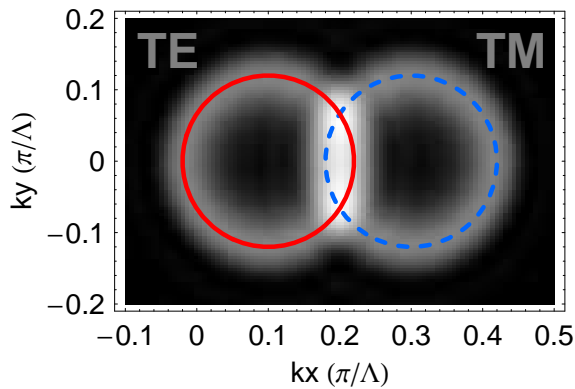}\\
\caption{Polarization-entangled photon pairs can be extracted from
the intersection of the TE and TM rings of the Type II emission
according to a well-known scheme~\cite{paul,pbs,paulpbs}.}
\label{fig:entang}
\end{figure}
To extract entangled photon pairs, we proceed as in the scheme for
the extraction of entangled photons from non-linear
crystals~\cite{paul}, by collecting photon pairs from the
intersection between the TE emission and the TM emission (see
figure~\ref{fig:entang}). At each intersection the photons will be
polarized TE(H) or TM(V) with equal amplitude. However since the
photons are emitted in TE/TM pairs, which ever the polarization of
a photon at one intersection, we know the  twin photon at
the other intersection will have opposite polarization.  Adding
the amplitudes we obtain a maximally entangled polarization state:
\begin{equation}
\frac{1}{\sqrt{2}} \big(\ket{H}\ket{V}+e^{i\Upsilon}\ket{V}\ket{H}\big)
\end{equation}
where $\Upsilon$ is a phase that can be easily tuned experimentally, for example by placing a birefringent element in one of the paths.

The existence of a fixed phase relation between the two terms
depends on the genuine impossibility of deducing the polarization of
a photon collected at one of the intersections from any of its
other properties. Such a coupling between degrees of freedom
reduces the coherence between the terms. There are many techniques
to recover the coherence, for example the use of compensating
crystals~\cite{paul} or of a polarizing
beam-splitter~\cite{paulpbs,pbs}.

\section{Conclusions}
\label{sec:concl}

We have presented a fully quantum-mechanical treatment of the down-conversion process in non-linear photonic crystals, showing how to calculate the emission analytically in one dimensional structures. We have applied the theory to a realistic  one-dimensional  structure that consists  of alternating layers of  Al$_{0.4}$Ga$_{0.6}$As and Air and demonstrated  that entangled photons with a wavelength of 1500nm can be extracted from the down-conversion emission that results from the decay of  pump photons having a wavelength of 750nm.

We now suggest some possible extensions of the present work.
The results for the $\chi^{(2)}$ mediated emission of section~\ref{sec:fff} are valid for structures of one, two and three dimensions. An obvious extension  is thus the explicit  analysis of Bloch-wave phase-matching in two  and three-dimensional structures. An advantage of the use of two-dimensional structures is that  two-dimensional structures with a large index contrast can be  more easily fabricated than one dimensional structures with a large index contrast.
Another interesting extension is the analysis of down-conversion in the case where the pump photons arrive in short pulses or wave-packets (see Ref.~\cite{walmsley} for a discussion of the problem in BBO). This brings up the topic of group velocity dispersion in photonic crystals which  can differ significantly from its counterpart in natural crystals. It seems likely that the possibility of tuning this form of dispersion will provide interesting  possibilities. Another interesting avenue of research might be to investigate the potential advantages of down-converting to frequencies near a photonic band edge, where the density of states is considerably larger than that at the centre of a band, thus further improving the efficiency. Finally it might be interesting to consider down-conversion of frequencies close to the non-linear  material absorption band-gap, where the non-linearity can be significantly enhanced.

In conclusion. the high degree of control afforded by the freedom of choosing the crystal geometry combined with the high non-linearity of semiconductor materials for which many advanced fabrication techniques have been developed make non-linear photonic crystals a promising  source of  entangled photon pairs.

\appendix


\section{Derivation of the Bloch-wave amplitudes}
\label{sec:appendix1}

In this Appendix we derive an expression for the Fourier coefficients of the periodic part of the electric field in one-dimensional Bragg structures. Throughout we will drop the TE(M) labels since the derivation is identical for the two cases. We begin by combining Equations~\ref{eq:g1} and~\ref{eq:g2} for the eigen-amplitudes of the forward and backward moving components of the electric field in each layer, with the expression for the electric field in the  nth layer (Eqns.~\ref{eq:ggg1} and~\ref{eq:ggg2}) to obtain:
\begin{eqnarray}
\mathbf{E}^{a}_{n}=e^{i(\omega t-\mathbf{K}\cdot\mathbf{r})}&\Big(& (a^+_0 \hat{\mathbf{\varepsilon}}^a_+)e^{-ik^a_z(z-n\Lambda )}e^{i\mathrm{K}_z(z-n\Lambda )}+\nonumber \\
& & (a^-_0 \hat{\mathbf{\varepsilon}}^a_-)e^{ik^a_z(z-n\Lambda)}e^{i\mathrm{K}_z(z-n\Lambda )}\Big)\label{eq:gg1} \\
\mathbf{E}^b_{n}=e^{i(\omega t-\mathbf{K}\cdot\mathbf{r})}&\Big(& (b^+_0 \hat{\mathbf{\varepsilon}}^b_+)e^{-ik^b_z(z-n\Lambda )}e^{i\mathrm{K}_z(z-n\Lambda )}+\nonumber \\
& & (b^-_0 \hat{\mathbf{\varepsilon}}^b_-)e^{ik^b_z(z-n\Lambda)}e^{i\mathrm{K}_z(z-n\Lambda )}\Big). \label{eq:gg2}
\end{eqnarray}
Comparing these relations with Eq.~\ref{eq:ef} then yields the following relation:
\begin{eqnarray}
\sum_\mathbf{n}\tilde{\bm{\varepsilon}}_{\mathbf{K},\lambda}\Big(n \frac{2 \pi}{\Lambda}\Big)e^{i n \frac{2 \pi}{\Lambda} z} & = & \sum_{\mathrm{layers}}E_{+,n}^a(z) \hat{\varepsilon}_+^a + E_{-,n}^a(z) \hat{\varepsilon}_-^a  \nonumber \\ &+& E_{+,n}^b(z) \hat{\varepsilon}_+^b + E_{-,n}^b(z) \hat{\varepsilon}_-^b
\label{eq:comp}
\end{eqnarray}
Where $E^{a(b)}_{+/-}(z)$ represent the electric field propagating in the forward/backward direction in material a(b). In the nth layer they are given by:
\begin{eqnarray}
E^a_{\pm,n}(z) &=& a^\pm_0 e^{\mp ik^a_z(z-n\Lambda)} e^{iK_z(z-n\Lambda)}\label{eq:gogol}\\
E^b_{\pm,n}(z) &=& b^\pm_0 e^{\mp ik^b_z(z-n\Lambda)} e^{iK_z(z-n\Lambda)}.\label{eq:gogol2}
\end{eqnarray}
To find the amplitudes of the of the Fourier coefficients $\tilde{\bm{\varepsilon}}_{\mathbf{K},\lambda}$  we evaluate the Fourier transform of both sides of Eq.~\ref{eq:comp}.
We define the Fourier transform $\tilde{f}$ of a function $f$ as:
\[
\tilde{f}(q)=\frac{1}{\sqrt{2 \pi}}\int_{-\infty}^\infty \mathrm{d}z f(z) e^{-i q z}
\]
The Fourier transform of the left hand side  of Eq.~\ref{eq:comp} is:
\begin{equation}
\sqrt{2 \pi}\sum_\mathbf{n}\tilde{\bm{\varepsilon}}_{\mathbf{K},\lambda}\Big(n \frac{2 \pi}{\Lambda}\Big)\delta\Big(q_z-n \frac{2 \pi}{\Lambda}\Big). \label{eq:ftlhs}
\end{equation}
To evaluate the Fourier transform of the right hand side  of Eq.~\ref{eq:comp} it is convenient to  first re-express the sum over layers in terms of the following:
the  top hat function $T_{x_1,x_2}(x)$:
\[
T_{x_1,x_2}(x)=\left\{ \begin{array}{ll} 1, & x_1 \leq  x  \leq x_2 \\ 0, &  x<x_1, x>x_2  \end{array}\right. ,
\]
with $x_2>x_1$, the comb function $\Pi_{(\Lambda, 2N+1)}(z)$:
\[
\Pi_{(\Lambda, 2N+1)}(z)=\sum_{n=-N}^N \delta(z-n\Lambda),
\]
and the convolution of functions $f$ and $g$: $f\star g$:
\[
[f\star g](x)=\int_{-\infty}^\infty \mathrm{d}z f(z) g(x-z).
\]
In terms of these functions we have:
\begin{eqnarray}
& \sum_{\mathrm{layers}}& E_{+,n}^a(z) \hat{\varepsilon}_+^a + E_{-,n}^a(z) \hat{\varepsilon}_-^a   + E_{+,n}^b(z) \hat{\varepsilon}_+^b + E_{-,n}^b(z) \hat{\varepsilon}_-^b  \\ &=&E_{+}^a(z) \hat{\varepsilon}_+^a + E_{-}^a(z) \hat{\varepsilon}_-^a  \nonumber  
 +E_{+}^b(z) \hat{\varepsilon}_+^b + E_{-}^b(z) \hat{\varepsilon}_-^b. \nonumber
\end{eqnarray}
with:
\begin{eqnarray}
E^a_\pm(z) &=& [(E^a_{\pm,n=0} T_{-a,0})\star\Pi_{(\Lambda,2N+1)}](z) \label{eq:e} \\
E^b_\pm(z) &=& [(E^b_{\pm,n=0} T_{-\Lambda,-a})\star\Pi_{(\Lambda,2N+1)}](z).\label{eq:ee}
\end{eqnarray}
$E^a_\pm(z)$ and $E^b_\pm(z) $ are well defined functions at all points in the the crystal. They are equal to 0 outside layers $a$ and $b$ respectively  and  equal to $E^a_{\pm,n}(z)$ (Eq.~\ref{eq:gogol}) and $E^b_{\pm,n}(z)$ (Eq.~\ref{eq:gogol2}) in the $n$th layer.
The Fourier transform of $E^a_\pm(z)$ and $E^b_\pm(z) $ can be simplified using the standard relation:
\begin{equation}
\widetilde{f \star g} = \tilde{f} \tilde{g}
\end{equation}
to obtain:
\begin{eqnarray}
\tilde{E}^a_\pm &=& (\widetilde{E^a_{\pm,n=0} T_{-a,0}}) \tilde{\Pi}_{(\Lambda,2N+1)} \nonumber \\
\tilde{E}^b_\pm &=& (\widetilde{E^b_{\pm,n=0} T_{-\Lambda,-a}}) \tilde{\Pi}_{(\Lambda,2N+1)}.
\end{eqnarray}
If we evaluate the Fourier transform of
 $\Pi_{(\Lambda, 2N+1)}$:
\begin{equation}
[\tilde{\Pi}_{(\Lambda, 2N+1)}](q_z) =\frac{1}{\sqrt{2\pi}}\frac{\sin\big[q_z \frac{\Lambda}{2}(2N+1)\big])}{\sin\big[q_z \frac{\Lambda}{2}\big]},
\end{equation}
and take the limit of $N\longrightarrow\infty$, we obtain:
\begin{equation}
\lim_{N\longrightarrow\infty} \tilde{\Pi}_{(\Lambda, 2N+1)}=\sqrt{2\pi}\Pi_{(\frac{2 \pi}{\Lambda},\infty)}
\end{equation}
comparison in this limit with the left hand side of Eq.~\ref{eq:ftlhs}  then leads to the following expression for $\tilde{\bm{\varepsilon}}_{\mathbf{K},\lambda}$:
\begin{eqnarray}
&& \tilde{\bm{\varepsilon}}_{\mathbf{K},\lambda}\Big(n \frac{2 \pi}{\Lambda}\Big) =\nonumber \\ 
&&\widetilde{E^a_{+,0} T_{-a,0}}\Big(n \frac{2 \pi}{\Lambda}\Big) \hat{\varepsilon}^a_+ + \widetilde{E^a_{-,0} T_{-a,0}}\Big(n \frac{2 \pi}{\Lambda}\Big) \hat{\varepsilon}^a_- \nonumber \\
&&+ \widetilde{E^b_{+,0} T_{-\Lambda,-a}}\Big(n \frac{2 \pi}{\Lambda}\Big) \hat{\varepsilon}^b_+ + \widetilde{E^b_{-,0} T_{-\Lambda,-a}}\Big(n \frac{2 \pi}{\Lambda}\Big) \hat{\varepsilon}^b_-.
\label{eq:coeffsa}
\end{eqnarray}
 Evaluating $\widetilde{E^a_{\pm,0} T_{-a,0}}(q_z)$ and
$\widetilde{E^b_{\pm,0} T_{-\Lambda,-a}}(q_z)$:
\begin{eqnarray}
&&\widetilde{E^a_{\pm,0} T_{-a,0}}(q_z)=\nonumber \\ 
&&\frac{a}{\sqrt{2 \pi}} a^\pm_0 \frac{\sin[(K_z\mp k^a_z-q_z)\frac{a}{2}]}{(K_z\mp k^a_z-q_z)\frac{a}{2}} e^{-i(K_z\mp k^a_z-q_z)\frac{a}{2}}\label{eq:last1}\\
&&\widetilde{E^b_{\pm,0} T_{-\Lambda,-a}}(q_z)=\nonumber \\
&& \frac{b}{\sqrt{2 \pi}} b^\pm_0 \frac{\sin[(K_z\mp k^b_z-q_z)\frac{b}{2}]}{(K_z\mp k^b_z-q_z)\frac{b}{2}} e^{-i(K_z\mp k^b_z-q_z)(a+\frac{b}{2})}\label{eq:last2}
\end{eqnarray}
  completes our derivation. Combining Eq.~\ref{eq:last1} and Eq.~\ref{eq:last2} with Eq.~\ref{eq:coeffsa},   we obtain the result given in section~\ref{sec:yariv}.

\begin{acknowledgements}
We acknowledge J.F.~Hodelin for useful discussions,  M.~Rakher and F.~Azhar for useful comments on the manuscript. This work was supported by NSF grant No. PHY-0304678 and DARPA grant No. MDA972-01-1-0027.
\end{acknowledgements}

\end{document}